\documentclass[a4paper,fleqn]{cas-sc}
\usepackage{lineno}

\newcommand*\patchAmsMathEnvironmentForLineno[1]{
  \expandafter\let\csname old#1\expandafter\endcsname\csname #1\endcsname
  \expandafter\let\csname oldend#1\expandafter\endcsname\csname end#1\endcsname
  \renewenvironment{#1}
     {\linenomath\csname old#1\endcsname}
     {\csname oldend#1\endcsname\endlinenomath}}
\newcommand*\patchBothAmsMathEnvironmentsForLineno[1]{
  \patchAmsMathEnvironmentForLineno{#1}
  \patchAmsMathEnvironmentForLineno{#1*}}
\AtBeginDocument{
\patchBothAmsMathEnvironmentsForLineno{equation}
\patchBothAmsMathEnvironmentsForLineno{align}
\patchBothAmsMathEnvironmentsForLineno{flalign}
\patchBothAmsMathEnvironmentsForLineno{alignat}
\patchBothAmsMathEnvironmentsForLineno{gather}
\patchBothAmsMathEnvironmentsForLineno{multline}
}
\usepackage[authoryear,longnamesfirst]{natbib}
\usepackage{physics}
\usepackage[version=4]{mhchem}
\usepackage{ulem}

\def\tsc#1{\csdef{#1}{\textsc{\lowercase{#1}}\xspace}}
\tsc{WGM}
\tsc{QE}

\begin{document}
%\linenumbers
\let\WriteBookmarks\relax
\def\floatpagepagefraction{1}
\def\textpagefraction{.001}

\shorttitle{Monte Carlo Simulation of Impact Vapor Plume Reactions}    

\shortauthors{Ochiai, Ida, Shoji}  

\title[mode = title]{Exploring Impact Vapor Plume Reactions from Asteroidal Impacts: Monte Carlo Simulations and Implications for Biomolecules Synthesis}

\author[1]{Yoko Ochiai}[orcid=0009-0007-9894-3590]
\ead{ochiai@elsi.jp}
\affiliation[1]{organization={Earth-Life Science Institute, Institute of Science Tokyo},
            addressline={2-12-1 Ookayama}, 
            city={Meguro-ku, Tokyo},
%          citysep={}, % Uncomment if no comma needed between city and postcode
            postcode={152-8550}, 
            %state={},
            country={Japan}}

\author[1]{Shigeru Ida}
% Email id of the second author
\ead{ida@elsi.jp}

\author[2]{Daigo Shoji}
% Email id of the second author
\ead{shohji.daigo@jaxa.jp}
\affiliation[2]{organization={Institute of Space and Astronautical Science, Japan Aerospace Exploration Agency},
            addressline={3-1-1 Yoshinodai}, 
            city={Chuo-ku, Sagamihara, Kanagawa},
%          citysep={}, % Uncomment if no comma needed between city and postcode
            postcode={252-5210}, 
            %state={},
            country={Japan}}

\begin{abstract}
During a hypervelocity impact, both the impactor and target materials evaporate, generating an impact vapor plume with temperatures reaching several thousand K.
As the plume cools through adiabatic expansion, chemical reactions are predicted to quench, leading to a non-equilibrium composition.
Previous experiments simulating meteorite impacts on the early Earth have reported the formation of biomolecules such as amino acids and nucleobases, suggesting that the chemical reactions within impact vapor plumes may have contributed to the origins of the building blocks of life. 
However, it is still unclear how chemical reactions proceed during the cooling impact vapor plume and lead to the synthesis of such organic molecules. 
In this study, to investigate the evolution of chemical composition within impact vapor plumes, we conducted a Monte Carlo chemical reaction simulation for complex organic synthesis, developed in our previous work (Ochiai, Y., Ida, S., Shoji, D., [2024], Astron. Astrophys., 687, A232).
In conventional kinetic model-based studies, chemical species and their associated reaction pathways are predefined to calculate the time evolution of chemical compositions using the thermodynamic data of these species and reaction rate coefficients.
In contrast, our model does not rely on a predefined reaction network; instead, it utilizes imposed conditions for chemical changes and an approximate method for calculating reaction rates suited to our objectives. 
Additionally, we developed a new approach to couple these chemical reaction calculations with the rapid temperature and pressure decay in the vapor plume.
Results show diverse organic molecule production depending on the impactor materials assumed in this study (LL, CI, and EL chondritic types). 
These products include important precursors to biomolecules such as amino acids, sugars, and nucleobases. 
On the other hand, for all impactor compositions, the abundance of biomolecules themselves remains extremely low throughout the reactions from an impact to quenching. 
Therefore, our results suggest that biomolecules are not directly produced in impact vapor plumes but rather synthesized through reactions of these precursor molecules in aqueous solutions, following H$_2$O condensation as the vapor plume cools.
Many of the detected organic compounds, including the precursor molecules such as imine compounds and formamide, are not included in the reaction networks of previous kinetic model simulations, and their formation has not been predicted.
This demonstrates the effectiveness of our Monte Carlo simulation as a powerful tool for investigating the synthesis of low-abundance organic compounds, including biomolecules.

\end{abstract}

\begin{highlights}
\item A Monte Carlo simulation we developed was applied to chemical reactions in impact vapor plumes
\item Compositional evolution of vapor plumes was calculated for three different impactor materials
\item The simulations demonstrated the synthesis of diverse organic molecules without predefined reaction pathways
\item Key precursors to biomolecules were identified among the synthesized organic molecules
\item Results suggest biomolecules are formed through liquid-phase reactions after H$_2$O condenses during plume cooling
\end{highlights}
% Keywords
\begin{keywords}
 Prebiotic chemistry \sep Organic chemistry \sep Impact processes \sep Earth \sep Asteroids
\end{keywords}

\maketitle

%\linenumbers
% Main text
\section{Introduction}\label{Introduction}

%"Prebiotic chemistry induced by impact events"
The origin of the building blocks of life remains a critical question in understanding the emergence of life. Organic synthesis triggered by meteorite impacts has been discussed as a potential mechanism for providing biomolecules to the early Earth \citep[e.g.,][]{Bar-Nun1970-xq, McKay1997-lf}. During hypervelocity impacts, the impactor materials are vaporized, forming a vapor plume rich in volatile elements such as carbon (C), nitrogen (N), oxygen (O), and hydrogen (H) \citep[e.g.,][]{Sugita2003-qb, Sugita2003-hu}.
Due to the extremely high temperature and pressure within a vapor plume, chemical reactions can proceed vigorously, potentially leading to the synthesis of organic molecules \citep[e.g.,][]{Mukhin1989-yh}. Previous studies have reported the formation of biomolecules, including amino acids and nucleobases, in shock-recovery experiments \citep{Furukawa2008-hx, Furukawa2015-xx, Takeuchi2020-yr} and in laser shock wave plasma experiments \citep{Ferus2017-he} simulating meteorite impacts on the early Earth.

Consequently, impact events in the early Earth, particularly during the Late Heavy Bombardment (LHB)
%\citep[e.g.,][]{Schoenberg2002-fy, Gomes2005-hk,Bottke2012-ne},lunar crater records suggest that impacts of extraterrestrial objects occurred with great frequency during those periods \citep{Culler2000-pf}
when the frequency of impacts was significantly high after the ocean was formed \citep[e.g.,][and references therein]{Bottke2017-zv, Lowe2018-fu}, may have played a crucial role in the emergence of life. 
However, the current understanding of chemical reactions within an impact vapor plume is limited, and the specific reaction pathways leading to the synthesis of complex biomolecules, as observed in experimental studies, remain unclear. %The essential factors and key processes regarding the shock synthesis remain unclear. 
Therefore, this study aims to elucidate the mechanisms of organic synthesis within impact vapor plumes, focusing on amino acids, one of the biomolecules identified in impact experiments.

%"Quenching process in vapor plumes"
The post-shock temperature of vapor plumes can reach extremely high values (>1000 K) as part of the impact energy is converted into the thermal energy \citep[e.g.,][]{Pierazzo1998-xz}. Following the impact, the vapor plume undergoes adiabatic expansion and rapid cooling on a timescale of a few seconds. Initially, the vapor plume attains thermodynamic equilibrium, as chemical reactions occur rapidly compared to the cooling rate. However, since reaction rates highly depends on the temperature, the timescale for chemical reactions increases significantly as the vapor cools. Consequently, chemical reactions within the vapor plume are quenched during the cooling.
Although quench temperatures depend on impact conditions and molecular species, they are generally predicted to fall within the range of $1000-3000$ K \citep[e.g.,][]{Gerasimov1999-mh, Fegley1986-vs, Ishimaru2010-di}.

%Previous study - kinetic model
%To date, the quenching compositions of impact vapor plumes have been estimated by some theoretical approaches.One approach assumes that quenching occurs at a specific temperature, with the resulting composition determined by the equilibrium state at that temperature \citep[e.g.][]{Kress2004-wy}. One approach is the time-scale model, which accounts for the differences in quenching temperatures among chemical species and assumes that the quenched abundance of each species is given by the equilibrium composition at its respective quenching temperature \citep[e.g.][]{}.
\citet{Ishimaru2010-di} investigated the quenching process of chemical reactions within a vapor plume using a chemical kinetic model, which calculates the temporal evolution of chemical species concentrations based on the reaction rates of elementary reactions. To account for both chemical reactions and the changes in temperature and pressure within the vapor plume, they coupled the chemical kinetic model with a one-dimensional radial hydrodynamic calculation, in which the expanding vapor plume was approximated as a gas hemisphere.
Their results demonstrated that not only the quenching temperatures vary between species, but the quenched abundances can also deviate--by a factor of 5 to 9 in the case of HCN--from those estimated by the equilibrium compositions of the temperature at which the timescale of decomposition of that species is equal to the cooling timescale.
This suggests that the quenched composition cannot be simply inferred from equilibrium composition, and that the chemical reactions within the vapor plume should be addressed using a kinetic approach. 
However, kinetic models require the prior specification of all the reactions included in the model, which can introduce biases into the reaction networks (set of chemical species and their associated elementary reactions).
The kinetic model (GRI-Mech version 3.0) used by \citet{Ishimaru2010-di} was originally developed to simulate combustion reactions of small hydrocarbons, such as methane and ethane, and may not account for potential products, including complex organic compounds, in impact vapor plumes. Therefore, to avoid excluding important potential reactions, an alternative approach that does not assume specific reaction pathways is needed.

%Previous study - AIMD simulations
Ab initio molecular dynamics (AIMD) simulations provide a method for simulating chemical reactions without the need for predefined reaction pathways and parameters, relying solely on quantum mechanical principles.
\citet{Goldman2010-np}, \citet{Goldman2013-jr} and \citet{Shimamura2019-ll} employed AIMD simulations to investigate post-shock reactions following the impacts of a comet and an Fe-bearing meteorite, respectively, with a focus on the synthesis of organic molecules. \citet{Goldman2010-np} demonstrated the formation of C-N bonded oligomers following compression, suggesting that complex organic molecules could be synthesized within vapor plumes.
However, the timescale applicable to chemical reactions in AIMD simulations is generally constrained to a few picoseconds due to the high computational costs involved. This limitation presents a challenge for achieving a comprehensive understanding of the quenching process, which occurs over a timescale of several seconds.

%Method in this study
Therefore, in this study, we employed the Monte Carlo simulation method for chemical reactions developed in our previous works \citep{Takehara2022-si, Ochiai2024}. 
In this method, a set of molecules is specified, and all possible reactions among them are listed. A single reaction is then selected based on probabilities weighted by reaction rates, and the molecular set is updated accordingly. By iterating this selection process, we perform a Monte Carlo simulation of chemical reactions that incorporates chemical kinetics without the need for predefined reaction networks.
This model significantly reduces computational costs by evaluating the reaction rates approximately, allowing for the exploration of the complete reaction process in a vapor plume until quenching occurs.
The adopted approximations are appropriate for our purpose, which is to obtain a comprehensive overview of chemical reactions leading to complex organic molecules.
Due to this approximation, some degree of inaccuracy in the reaction rates is inevitable, making this method less suitable for determining the precise concentrations of individual chemical species in the system. However, because this model considers all potential reaction pathways with relatively low computational cost, it is well-suited for exploring the synthesis of organic molecules involving complex reaction pathways that are difficult to capture with conventional kinetic models.
Notably, our simulation results showed the synthesis of key precursors to biomolecules in the impact vapor plume, which were not included in the reaction network model used by the previous study \citep{Ishimaru2010-di}, as shown in \Secref{sec:amionacid}.

The original model in \citet{Ochiai2024} was primarily developed to simulate surface reactions of icy dust particles driven by UV irradiation in protoplanetary disk environments ($\sim 50-100\ \rm K$). Hence, several modifications were made to adapt the model for our current study. First, we introduced the consideration of entropy changes in reactions to accommodate gas-phase reactions, which occur at higher temperatures than ice surface reactions. Additionally, we developed a method to couple the Monte Carlo simulations with temporal changes in temperature and pressure, enabling the model to simultaneously track the chemical reactions and the rapid temperature and pressure change occurring within the vapor plume over time.
Details of these updates are provided in \Secref{sec:reactionrates}.

%Summary of this paper
In \Secref{Method}, we describe the fundamental methodology of the Monte Carlo simulation developed by \citet{Ochiai2024}, along with the updates applied in this study. We also explain the procedures for calculating temperature and pressure variations within an impact vapor plume and how these calculations are integrated with the chemical reaction simulations.
Additionally, we detail the initial conditions for the simulations, including the initial temperature, pressure, and composition of the impact vapor plumes.
In \Secref{Result}, we present the compositional evolution of vapor plumes for three different impactor materials, focusing on the mole fraction changes of major chemical species and organic molecules.
Our results show the synthesis of a wide variety of organic molecules within impact vapor plumes, including species not accounted for in the reaction network used in previous study. Furthermore, some of these organic products may play crucial roles in subsequent biomolecule synthesis.
In \Secref{Discussion}, we discuss the potential influence of the approximations used in estimating reaction rates and the omission of rock-forming elements in impact vapor plumes in this study.
In \Secref{Conclusion}, we summarize our results.

\section{Method} \label{Method}

In this section, we briefly explain a numerical method of the Monte Carlo simulation proposed in \citet{Ochiai2024} and the modifications newly made  in this study to account for asteroid impacts.
The basic scheme of the Monte Carlo simulation presented in \Secref{sec:montecarlo} follows directly the one employed in \citet{Ochiai2024}.
However, the underlying environmental conditions represented by temperature $T$ and pressure $P$ differ from the previous model %\citet{Ochiai2024} 
(see \Secref{sec:TPchange} and \Secref{sec:coupling}).
Thus, the weight of probabilities to choose a next reaction from the reaction candidates, explained in Section~\ref{sec:reactionrates}, was updated to a more general version than that in \citet{Ochiai2024} to apply the simulation for higher-temperature environments such as asteroid impacts, as well as the low temperature environments such as icy dust surface exposed to UV irradiation. 
We emphasize that the strength of our simulation model lies in its versatility: it can be applied to the synthesis of organic molecules in diverse environments and under different physical and chemical processes, simply by adjusting the environmental conditions while maintaining the same fundamental scheme.

\subsection{Basic Scheme of the Monte Carlo Simulation: Tracking Chemical Reaction Sequences}\label{sec:montecarlo}

\begin{figure}
\centering
\includegraphics[width=0.6\linewidth]{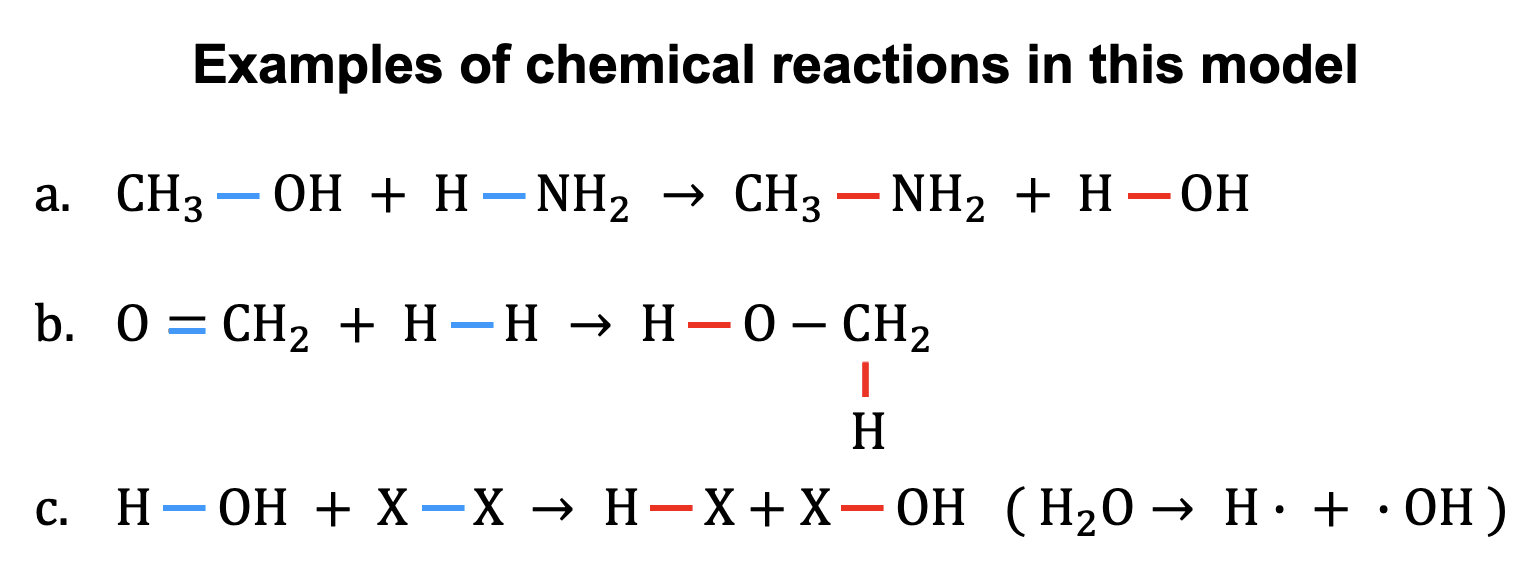}
\caption{
Examples of chemical reactions considered in this model. In each reaction, bonds being broken are colored blue, and bonds being formed are colored red. As shown in the second reaction, double bonds (and triple bonds) are treated as two (or three) independent bonds. The atom $\rm X$ shown in the third reaction is a hypothetical element used to represent radical species within our reaction scheme (see \Secref{sec:montecarlo}).
}
\label{fig:ex_reaction}
\end{figure} 

The fundamental unit of this Monte Carlo simulation is a molecular set consisting of several dozen molecules and/or radicals.
Every chemical species within a molecular set are characterized by its constituent atoms and the bonds between these atoms.
The initial step in a chemical reaction simulation is to list the all chemical reaction candidates that are possible to occur within the molecular set. 
However, the reactions listed here are limited to the exchange of two bonds, specifically the breaking and recombination of two bonds, which represent the minimal components of chemical reactions.
Examples of chemical reactions considered in this simulation are shown in \Figref{fig:ex_reaction}.
Next, a single reaction is randomly selected from the candidates based on the weighted probabilities evaluated from reaction rates (see Section \ref{sec:reactionrates}).
Then, the molecular set after the selected chemical reaction occurs is designated as a new starting set.
We refer to this entire series of calculations up to this point as a single reaction step.

For a given initial condition, the reaction calculation up to a specific reaction step number is performed multiple times (typically $10^3$ trials in this study) using different random number seeds to ensure sufficient sampling.
Consequently, the statistical molecular evolution in the molecular set is obtained.

To represent radical reactions, we use a hypothetical molecule, $\rm X_2$, following \citet{Ochiai2024}.
$\rm X_2$ is not a physically existing molecule, but an element used solely for computational purposes.
Using a water molecule as an example, a radical formation reaction can be written as \Figref{fig:ex_reaction}c.
The products, H-X and X-OH, represent an H radical and an OH radical, respectively. 
The bond to an X atom indicates the presence of an unpaired electron on the binding atom, and its bond energy is therefore set to be zero. 
By introducing $\rm X_2$, radical reactions, which do not physically involve the exchange of two bonds, are still formally treated as chemical reactions within this model, defined as the exchange of two bonds.
In \citet{Ochiai2024}, the X-X bond energy was assigned a negative value to represent the photon energy associated with UV irradiation in a protoplanetary disk. Since photon irradiation is not considered in this study, the bond energy of $\rm X_2$ is set to zero.

\subsection{Probability weighting and estimation of reaction rates} \label{sec:reactionrates}

In this section, we explain how the probabilities for the reaction candidates are weighted during the chemical reaction calculations described earlier. 
In this study, we adopted a different form of the weight from the one used in the previous model by \citet{Ochiai2024}. 

\begin{table}[hbtp]
  \centering
  \begin{tabular}{|lr|lr|lr|lr|lr|lr|}
    \hline
     \multicolumn{6}{|c|}{Bond energy [kJ/mol]} \\
    \hline \hline
    C $-$ C  & 347 & N $-$ N  & 161 & O $-$ O  & 146 \\
    C $=$ C  & 611 &  N $=$ N  & 456 & O $=$ O  & 498 \\
    C $\equiv$ C & 837 & N $\equiv$ N  & 946 & & \\ \hline
    C $-$ N  & 305 &   C $-$ O  & 358 & N $-$ O  & 230 \\
    C $=$ N  & 615 & 
    C $=$ O  & 745$^{*}$ & N $=$ O  & 598 \\
    C $\equiv$ N  & 891 & C $\equiv$ O & 1072 & & \\ \hline
    C $-$ H  & 414 & N $-$ H  & 389 &
    O $-$ H  & 464 \\
    H $-$ H  & 436 & & & &
    \\
    \hline
  \end{tabular}
  \caption{Bond energies used in this study \citep{sanderson}. 
  (*) For $\rm{CO_2}$, the C=O double bond energy is 799 kJ/mol instead of 745 kJ/mol.
  }
    \label{table:bond}
\end{table}

In \citet{Ochiai2024}, the probability weighting for each reaction, $W$, was given as follows:
%\brown{\citet{Ochiai2024} considered relatively low temperature environments, $T=50-100\,\rm K$, and adopted the probability weighting for each reaction, $W$, given by}
\begin{equation}
    W=\exp\left(-\frac{E_{\rm a}}{RT}\right),
    \label{eq:W_lowT}
\end{equation}
where $E_{\rm a}$ is the activation energy of the reaction.
%which was approximately estimated from the enthalpy change of the reaction.
%the time scales on which the reactions proceeded and the difference between a first order reaction and a second order reaction.
This activation energy in Eq.~(\ref{eq:W_lowT}) was evaluated using the Bell–Evans–Polanyi principle:
\begin{align}
E_{\rm a} = \alpha' \Delta H + \beta',
\label{eq:BEP}
\end{align}
where $\Delta H$ is the enthalpy change of the reaction, and $\alpha'$ and $\beta'$ are empirical parameters.
According to density functional theory (DFT) calculations \citep{michaelides_2003,Wang_2011_evans,sutton_2012}, $\alpha' =1$ and $\beta' = 100 \, \rm kJ/mol$ were adopted for fiducial values in \citet{Ochiai2024}.
The enthalpy change $\Delta H$ was approximately calculated in the same manner as this study (see below).
%The enthalpy $H$ of molecules before and after the reaction is calculated by simply summing up bond energies of individual chemical bonds in the molecules  (Table.~\ref{table:bond}). With these approximations, we avoided computationally expensive quantum chemistry calculation to significantly speed up the simulation. 
This Arrhenius-type weight reflected the relative magnitude of the reaction rates among the candidate reactions, but it did not explicitly account for reaction timescales and differences between first- and second-order reactions.
Furthermore, since \citet{Ochiai2024} focused on systems where non-equilibrium states are maintained through reactions, consistency with thermodynamic equilibrium was not considered, which is important for high-temperature chemistry in an impact vapor plume.

Thus, in this study, while following  the framework that incorporates individual reaction rates into the simulation by weighting the probability of each reaction, we apply a modified weighting formula adjusted to the conditions of the impact vapor plume.
As explained in \Secref{sec:TPchange}, we incorporate the time evolution of temperature and pressure within the vapor plume into chemical reaction calculations. 
To achieve this, the timescales of chemical reactions are required, necessitating the estimation of the absolute values of reaction rates.
Therefore, in this study, we employ the Eyring equation to derive the reaction rates and directly use them to weight the probabilities instead of the previous weight equation (\ref{eq:W_lowT}).
The Eyring equation is based on the the transition state theory, which assumes the quasi-equilibrium between reactants and activated transition state complexes, and is written as follows:
\begin{equation}
    k_{\rm first}=\frac{\kappa k_{\rm B} T}{h} \exp\left(-\frac{\Delta^\ddag G^{\circ}}{RT}\right),
\label{eq:Eyring1}
\end{equation}
\begin{equation}
    k_{\rm second}=\frac{\kappa k_{\rm B} T}{h} \left( \frac{RT}{P^{\circ}} \right) \exp\left(-\frac{\Delta^\ddag G^{\circ}}{RT}\right),
%=\frac{\kappa k_{\rm B} T}{h} \exp{\left(\frac{\Delta S^\ddag}{R}\right)}\exp{\left(-\frac{\Delta H^\ddag}{RT}\right)},
\label{eq:Eyring2}
\end{equation}
where $k_{\rm first}$ and $k_{\rm second}$ are the rate constants of first-order reactions and second order reactions, respectively,  $\kappa$ is the transmission coefficient, $k_{\rm B}$ is the Boltzmann constant, $R$ is the gas constant, $T$ is the temperature, $h$ is the Planck constant, $P^{\circ}$ is the standard pressure (1 bar), and $\Delta^\ddag G^{\circ}$ is the standard Gibbs energy of activation (the difference in Gibbs energies between reactants and activated transition state complexes in their standard states).
We assume that the $\kappa = 1$ in this study.

To calculate $\Delta^\ddag G^{\circ}$, we assume the linear relationship between the standard Gibbs energy of activation $\Delta^\ddag G^{\circ}$ and the standard reaction Gibbs energy  $\Delta_r G^{\circ}$ (the difference in Gibbs energies between reactants and products in their standard states), as follows:
\begin{equation}
    \Delta^\ddag G^{\circ}= \alpha \Delta_r G^{\circ} + \beta,
\label{eq:LFER}
\end{equation}
where $\alpha$ and $\beta$ are the constants specified by the reaction type.
Although such linear relationships have been identified in many reactions in previous studies using quantum chemical calculations \citep[e.g.,][]{Salamone2021-aq, Yang2024-dg}, how  $\alpha$ and $\beta$ are determined for a given reaction has not yet been formulated.
Thus, in this study, we assume $\alpha =0.5$ and $\beta$ in the range of  $250 - 350\ \rm kJ/mol$, with $\beta=300\ \rm kJ/mol$ 
as the fiducial value, based on comparisons between our calculations and the results of \citet{Ishimaru2010-di} (see Appendix A for a detailed explanation of assumptions $\alpha$ and $\beta$).

The standard reaction Gibbs energy $\Delta_r G^{\circ}$ is expressed in terms of the standard reaction enthalpy $\Delta_r H^{\circ}$, temperature $T$, and the standard reaction entropy $\Delta_r S^{\circ}$, as follows:
\begin{equation}
    \Delta_r G^{\circ}= \Delta_r H^{\circ} - T \Delta_r S^{\circ}.
\label{eq:DeltaG}
\end{equation}
Since $\Delta_r H^{\circ}\sim O(100) \, \rm kJ/mol$ and $\Delta_r S^{\circ} \sim O(0.1) \, \rm kJ/mol$ (see Tables~\ref{table:bond} and \ref{table:entropy}), 
the entropy term is negligible for $T\sim 50-100\, \rm K$, as considered by \citet{Ochiai2024}.
However, because the temperature in an impact vapor plume generated by an asteroid impact can exceed $1000 \, \rm K$, accounting for entropy changes becomes essential.
Therefore, in this study, $\Delta_r H^{\circ}$ and $\Delta_r S^{\circ}$ are calculated using the methods described below.

As in the previous model, $\Delta_r H^{\circ}$ is given by the difference in the total bond energies (sum of the bond energies of the individual bonds that constitute a molecule) of reactants and products of each reaction.
Bond energies used in this study are shown in Table~\ref{table:bond}.

For the calculation of $\Delta_r S^{\circ}$, we approximate entropy
by considering only the contribution from translational motion, as follows \citep{Atkins}:
\begin{align}
    \frac{S_{\rm m}}{R}&\simeq 
    \frac{2}{3}+\frac{3}{2}\ln \left(\frac{2\pi M \, k_{\rm B} T}{h^2}\right) - \ln \left(\frac{P}{k_{\rm{B}} T}\right)+1 \notag\\
    & \simeq \ \frac{3}{2}\ln \left(\frac{M}{1\ \rm amu}\right) +\frac{5}{2}\ln \left(\frac{T}{300 \ \rm K}\right) - \ln \left(\frac{P}{1\ \rm bar}\right)+13 ,
\label{eq:entropy}
\end{align}
where $M$ is mass of the molecule, and the subscript ``m" denotes a value per mol.
The standard entropy and enthalpy of several molecules are shown in Table~\ref{table:entropy}.
\begin{table}[hbtp]
\centering
\begin{tabular}{|c||c|c|c|c|}\hline
     &  CH$_4$ &  H$_2$O   & CO$_2$ & H$_2$ \\ \hline\hline
$S^\circ$ [kJ/mol K]   & 0.143     & 0.144     & 0.156    & 0.117  \\  \hline
$H^\circ$ [kJ/mol]     &  1656     & 928           & 1598      & 436      \\  \hline        
\end{tabular}
  \caption{The standard entropy $S^\circ$ (at $T=298\,\rm K$ and $P = 1\,\rm bar$) for CH$_4$, H$_2$O, CO$_2$, and H$_2$, calculated by Eq.~(\ref{eq:entropy}).
  For comparison, the total bond enthalpy $H^\circ$, or the standard enthalpy change for atomization (e.g., \ce{H2O -> 2 H + O}), of these species is also provided, calculated using bond energies from Table\ref{table:bond}.
  }
  \label{table:entropy}
\end{table}

Setting the standard pressure $P^{\circ}$ to $1\ \rm bar$, $\Delta_{\rm r} S^{\circ}$ is represented as follows;
\begin{align}
    \frac{\Delta_{\rm r} S^{\circ}}{R}&=\sum_{J} \nu_J \frac{S^{\circ}_{\rm m} (J)}{R}\notag\\
    &\simeq \ \frac{3}{2}\ln\left(\prod_J M_J^ {\nu_J}\right) +\left(\sum_J \nu_J\right) \left[\frac{5}{2}\ln\left( \frac{T}{300 \ \rm K}\right)+13\right].
\label{eq:DeltaS}
\end{align}
where $J$ denotes the chemical species and $\nu_J$ is the corresponding stoichiometric number.
Since the dependence of molecular mass $M$, $P$, and $T$ are weak, $\Delta_{\rm r} S^{\circ}$ is generally dominated by the change in total number of molecules.

To summarize, in this study, the weight of probabilities to select the reactions is expressed as follows:
\begin{equation}
\renewcommand{\arraystretch}{2}
W=
\left\{
\begin{array}{ll}
{\displaystyle \frac{k_{\rm B} T}{h} \exp\left(\frac{\alpha \Delta_{\rm r} S^{\circ}}{R}\right)\exp\left(-\frac{\alpha \Delta_{\rm r} H^{\circ} + \beta}{RT}\right)} \lbrack R _1\rbrack & \text{: first-order reaction}\\
{\displaystyle \frac{k_{\rm B} T}{h} \left( \frac{RT}{P^{\circ}} \right) \exp\left(\frac{\alpha \Delta_{\rm r} S^{\circ}}{R}\right)\exp\left(-\frac{\alpha \Delta_{\rm r} H^{\circ} + \beta}{RT}\right)} \lbrack R_1 \rbrack \lbrack R_2 \rbrack& \text{: second-order reaction}
\end{array}
\label{eq:weight}
\right. ,
\end{equation}
where $[R_1]$ and $[R_2]$ are the molar concentrations of the reactants, derived using Eq.\eqref{eq:concentration}.
The enthalpy change is characterized by bond energy, while the entropy change is determined by the stoichiometric numbers, masses of the reactants and products, and temperature, as explained above.
By incorporating entropy into our model, it has been refined into a more general thermodynamics-based framework, allowing its application to higher-temperature reactions.
We discuss the impact of the approximations of reaction rates in \Secref{sec:approx}.

It should be noted that three-body reactions (e.g., \ce{H. + H. + M} (third body) \ce{<=> H2 + M}), which require an additional molecule or atom to either remove excess energy (forward direction in the reaction above) or supply the energy needed for bond dissociation (reverse direction), are not explicitly expressed in the current model. 
Since the third body does not undergo any change during the reaction, we treat these reactions without considering the third body in our simulations (i.e., the above reaction is treated as \ce{H. + H. <=> H2}).
However, the reaction rates of such reactions depend on the concentration of the third body, which our current model does not account for. Under the high-pressure conditions considered in this study ($10^7-10^{11}$ Pa), the concentration of third bodies is expected to be sufficiently high that the reaction rates become effectively independent of their concentration.
Furthermore, since the equilibrium composition does not depend on the reaction rates including those of three-body reactions, but is determined by the Gibbs energies of the molecular species, the dependence of reaction rates on third-body concentrations can be neglected under the high-temperature and high-pressure conditions of vapor plumes where equilibrium is achieved.
The kinetic model (GRI-Mech 3.0), which we compared our results with in \Secref{sec:comparison}, does include this concentration dependence in its calculations. Our model, despite neglecting this factor, remains consistent with the kinetic model (see \Secref{sec:comparison}).

\subsection{Temperature and pressure change of a vapor plume} \label{sec:TPchange}

In \citet{Ochiai2024}, a constant temperature $(\sim 50-100\, \rm K)$ was assumed throughout the simulations, with a transition from a UV-irradiated phase to a phase without UV irradiation, representing the sedimentation of an ice dust particle from the upper layer of the protoplanetary disk to the UV-shielded midplane. 
In contrast, in the case of an asteroid impact, both temperature $T$ and pressure $P$ are extremely high immediately after the impact, rapidly decreasing as the vapor plume undergoes adiabatic expansion.

This section presents the model for the decay of temperature and pressure ($T$ and $P$) in the impact vapor plume, which are essential parameters for our chemical reaction simulations. 
In this study, we adopt a fiducial impact case with an impact radius of $0.5\ \rm{km}$ and an impact velocity of $10\ \rm{km/s}$.
When an impact occurs, a portion of the impact energy is converted into thermal energy, resulting in a temperature increase.
The specific impact energy with the impact velocity $v$ is given by:
\begin{equation}
    E_{\rm{imp}} = \frac{1}{2} v^2 
    = 50\left(\frac{v}{10\ \rm{km/s}}\right)^2\ \rm{km}^2/\rm{s}^2.
\label{eq:E_imp}
\end{equation}
The specific thermal energy is represented as $c_P T$, where $c_P$ is specific heat of the substances.
With the heat conversion rate $\epsilon$, the post-impact temperature rise is expressed as:
\begin{equation}
    T_0 \sim \frac{\epsilon E_{\rm{imp}}}{c_P} 
    \sim 5 \times 10^3 \left(\frac{\epsilon}{0.1}\right)\left(\frac{v}{10\ \rm{km/s}}\right)^2 \left(\frac{c_P}{10^3\ \rm J/kg\cdot  K}\right)^{-1} \ \rm K.
\label{eq:post_impT}
\end{equation}
To fit the temperature estimated from \Eqref{eq:post_impT} with the results of \citet{Marchi2013-up}, which showed post-impact temperatures using the two-dimensional hydrocode; impact-Simplified Arbitrary Lagrangian Eulerian (iSALE), we adopt $\epsilon = 0.1$.
Consequently, the initial temperature of a vapor plume is set to $5000 \ \rm K$ in this study.
%Since the temperature derived from \Eqref{eq:post_impT} represents the one at the impact point, the initial temperature of a vapor plume is set to $5000 \ \rm K$.
%\brown{However, the choices of the initial $T$ and $P$ hardly affect the final results. As long as the initial $T$ and $P$ are high enough that the molecule abundances quickly establish the equilibrium at those $T$ and $P$, the final molecule abundances are regulated by quenching process during the cooling.}

To estimate the initial pressure of the impact vapor plume $P_0$, we use Hugoniot relation, as follows:
\begin{equation}
    P_0 \simeq \rho U_{\rm s} u_{\rm p} 
    \simeq \rho (C_{\rm imp} + S_{\rm imp} u_{\rm p} ) u_{\rm p}, 
\label{eq:post_impP}
\end{equation}
where $\rho$ is pre-shock density of the impactor, $U_{\rm s}$ is the shock-wave velocity, $u_{\rm p} $ is the particle velocity, $C_{\rm imp}$ is the bulk sound velocity of the impactor material, and $S_{\rm imp}$ is a material constant.
%We emphasize that $P_0$ is the pressure after the shock has passed through the impactor, but not the pre-shock pressure.
For simplicity, assuming the impactor and target are made of the same material, the particle velocity of the impactor is expressed as $u_{\rm p} =(1/2)\,v$ as the impactor and the target have the same pressure and the same velocity at the boundary surface after the impact.
Furthermore, assuming basalt as the material, and using the values for basalt presented in \citet{Sekine2008-ga}, $\rho = 2.70 \ \rm g/cm^3$ $C_{\rm imp}=3.5\ \rm km/s$ and $S_{\rm imp}=1.3$, we obtain $P_0 \simeq 140$ GPa.

Next, we derive the time variations of temperature and pressure in an impact vapor plume that adiabatically expands.
Although a realistic vapor plume is a hemisphere, we estimate the $T$-$P$ variations assuming the spherically symmetric vapor plume, for simplicity.
%to be spherically symmetric with respect to the impact point.
Given the expansion velocity (radial velocity) of the vapor plume as the sound speed, 
the radius of the vapor plume $r_{\rm vp}$ evolves as follows:
\begin{equation}
    \frac{dr_{\rm vp}}{dt}=c_s 
    \simeq 1.7 \left(\frac{T}{10^4 \ \rm K}\right)^{1/2}\left(\frac{\mu}{30}\right)^{-1/2}\ \rm km/s,
\end{equation}
where $\mu$ is the mean molecular weight of the vapor.
Using the normalized quantities; $\tilde{r}_{\rm vp} = r_{\rm vp}/r_{\rm imp}$ and $\tilde{T}=T/T_0$, 
where $r_{\rm imp}$ is the impactor radius (approximately the initial vapor plume's radius) and $T_0$ is the initial vapor plume temperature, the above equation is rewritten as follows:
\begin{equation}
    \frac{d\tilde{r}_{\rm vp}}{dt}\simeq C \tilde{T}^{1/2} \ ; \
\label{eq:dr/dt}
%\end{equation}
%where
% \begin{equation}
    C=3.3 \left(\frac{r_{\rm imp}}{0.5 \ \rm km}\right)^{-1}\left(\frac{T_0}{10^4 \ \rm K}\right)^{1/2}\left(\frac{\mu}{30}\right)^{-1/2}.
\end{equation}
Assuming adiabatic expansion of ideal gas, $\tilde{r}_{\rm vp}\simeq \tilde{T}^{\,-1/3(\gamma -1)}$, where $\gamma$ is the adiabatic exponent.
Substituting the this relation into \Eqref{eq:dr/dt}, 
\begin{equation}
    \frac{d\tilde{T}}{dt}\simeq -3(\gamma -1)C \tilde{T}^{\frac{1}{3(\gamma -1)}+\frac{3}{2}}.
\label{eq:dT/dt}
\end{equation}
By solving this equation, we obtain the temperature variation, and the pressure variation can be determined using the relation between $\tilde{T}$ and $\tilde{P}(=P/P_0)$, as follows:
\begin{equation}
\tilde{P}= \tilde{T}^{\,\gamma/(\gamma -1)}.
\label{eq:TP_relation} 
\end{equation}
In this study, we use $\gamma = 1.4$ as a constant value.

\subsection{Coupling method}
\label{sec:coupling}

Based on \Eqref{eq:dT/dt}, the cooling time scale over which the temperature changes by $5\%$ is calculated as 
\begin{equation}
    t_{\rm cool}=\left|\frac{0.05 \tilde{T}}{d\tilde{T}/dt}\right|=\frac{0.05}{3(\gamma -1)C}\tilde{T}^{-\frac{1}{3(\gamma -1)}-\frac{1}{2}}\ \rm s.
\label{eq:t_cool}
\end{equation}
For $\gamma \sim 1.4$ and $\mu \sim 30$, 
\begin{equation}
t_{\rm cool} \sim 0.01 \, 
\left(\frac{r_{\rm imp}}{0.5 \, {\rm km}}\right) 
\,\left(\frac{T}{10^4 \, \rm K}\right)^{-1/2}
\ \rm s.
\end{equation}

On the other hand, the time step of the $i$-th reaction step in the Monte Carlo calculation, $\Delta t_i$, is expressed as
\begin{equation}
    \Delta t_i =\frac{1}{\sum k_{\rm{first},i}+ C_i\sum k_{\rm{second},i}} \ \rm s,
\label{eq:Delta_t}
\end{equation}
where %$k_{\rm{first},i}$ and $k_{\rm{second},i}$ are the rate constants of first-order reactions and second-order reactions at the $i$-th step, respectively. 
$C_i$ is the concentration calculated as 
\begin{equation}
    C_i = \frac{P_i/RT_i}{N_{\rm{molecules},i}},
\label{eq:concentration}
\end{equation}
and $N_{\rm{molecules},i}$ is the total number of molecules in the molecular set at the $i$-th step (see Appendix B for more details), and $\sum$ indicates the sum of $k_{\rm{first},i}$ and $k_{\rm{second},i}$ for all reaction candidates at the $i$-th step.
%Thus, $\Delta t_i$ is regulated by the fastest reaction among the candidates.
%The use of summation of $k$ is to avoid an artificially quench of the reaction sequence by a low-probability choice of relatively slow reaction.

\begin{figure}
\centering
\includegraphics[width=0.9\linewidth]{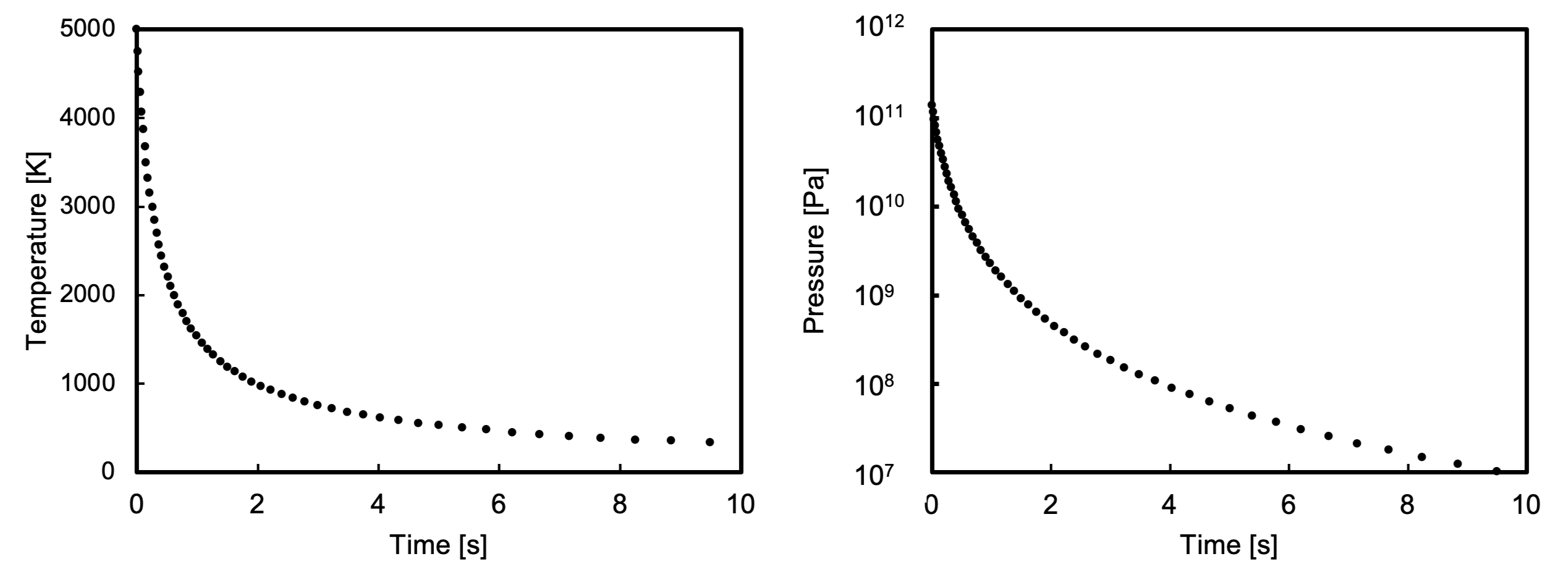}
\caption{Typical time evolution of temperature and pressure in impact vapor plume assumed in this study. Initial temperature and pressure are 5000 K and 140 GPa, respectively.
Each point represents a 5\% change in temperature calculated from \Eqref{eq:dT/dt}
and \Eqref{eq:TP_relation}, using $C = 2.3$ and $\gamma=1.4$.}
\label{fig:TPrelation}
\end{figure}

At each step of the reaction calculation, the weights of reaction probabilities (\Eqref{eq:weight}) are calculated using the input  temperature and pressure, and the cumulative reaction time is obtained by summing $\Delta t_i$.
In this study, to couple the reaction calculation with the temperature and pressure changes, the temperature and pressure are updated when the elapsed reaction time since the last temperature and pressure update exceeds $t_{\rm cool}$.
When the update occurs, $T$ is reduced by $5\%$, and $P$ is recalculated using \Eqref{eq:TP_relation} with the new $T$.
At the same time, $t_{\rm cool}$ is updated using the new $T$ and the mean molecular weight $\mu$ derived from the molecular set at that step (\Eqref{eq:t_cool}). 
Consequently, the chemical reaction calculations utilize step-wise $T$-$P$ functions, with temperature reduced by 5\% increments and the corresponding decrease of pressure.
Figure~\ref{fig:TPrelation} shows the typical time evolution of temperature and pressure in an impact vapor plume used in the current calculations.

In the initial high-temperature phase, since $\Delta t_i$ is extremely small, the system reaches equilibrium in a time much shorter than $t_{\rm cool}$.
Therefore, to reduce computational cost, the maximum number of reaction steps at a given temperature is set to $10^5$ steps, which is large enough for the current simulation set to reach equilibrium. Once this number is reached, the simulation transitions to the next temperature state.

%Since in the early high-temperature phase, the equilibrium of compositions is quickly established. If the condition that ?? is satisfied, we move onto the next $T$ phase, regarding that the compositions do not change until the end of this $T$ phase.

\subsection{Initial conditions}\label{sec:initialcondition}

For impactor materials, we assume three types of chondrites; carbonaceous, ordinary, and enstatite chondrites.
We calculate chemical reactions among only volatiles (C, H, N, and O species), assuming specific elemental compositions to each material.
Under high temperatures immediately after an impact, vaporization of rock components also occurs and they condense out during cooling.
The condensation of these components can trap volatile elements, possibly leading to the change in elemental compositions of the gas phase over time. 
However, the kinetics of these processes are not clear at the moment.
Thus, we ignore this effect based on the assumption that the condensation mostly completes at sufficiently high temperature.

\citet{Schaefer2010-an} focused on impact outgassing and calculated the equilibrium compositions in gas phase from chondrite compositions.
We determine the elemental compositions of our calculations based on the equilibrium gas composition of CI (carbonaceous chondrites), LL (ordinary chondrites), and EL (enstatite chondrites) chondrites at 1500 K and 100 bar calclulated in \citet{Schaefer2010-an}.
While this temperature and pressure condition is different from the initial condition assumed in this study, they showed that the compositions of the nominal condition (1500 K and 100 bar) are still valid at other temperatures and pressures.
These compositions also contain trace amounts of sulfur; however, sulfur is ignored in this study as its abundance is sufficiently small compared to C, N, O, and H.
Furthermore, in this study, mixing with nitrogen atmosphere was assumed as in previous experiments and the nitrogen ratio was increased from the equilibrium compositions in \citet{Schaefer2010-an}.
Although it is currently unclear whether and how an impact vapor plume mixes with the surrounding atmosphere, we assumed that $5 \%$ of total impact vapor volume come from pure $N_2$ atmosphere, representing the early Earth's atmosphere.
The elemental compositions and the corresponding initial sets of molecules used in this study are shown in Table \ref{table:composition}.

\begin{table}
    \centering
    \begin{tabular}{|c|c|c|c|} 
    \hline 
     type & atomic composition & C/H & O/H \\ \hline \hline
     LL (ordinary chondrite) & 50 H, 12 C, 4 N, 20 O & 0.24 & 0.40 \\ 
     CI (carbonaceous chondrite) & 50 H, 7 C, 4 N, 37 O & 0.14 & 0.74 \\ 
     EL (enstatite chondrite) & 12 H, 22 C, 4 N, 25 O & 1.83 & 2.08\\ 
     comet & 42 H, 12 C, 2 N, 21 O & 0.29 & 0.50 \\ 
    \hline
    \end{tabular}
    \caption{Atomic compositions for CI, LL, and EL chondrite types. 
    The comet type is used for the comparison with \citet{Ishimaru2010-di} (see \Secref{sec:comparison}).
    The number of atoms represents those included in a single molecular set of each simulation. These ratios between H, C, N, and O are determined based on the equilibrium compositions at 1500 K and 100 bar derived in \citet{Schaefer2010-an}.
    We assumed mixing of 5\% by volume of the ambient $N_2$ atmosphere into the vapor plume and increased the N fraction.
    }
    \label{table:composition}
\end{table}

Initial temperature and pressure of the impact vapor plume are 5000 K and 140 GPa, respectively as described in \Secref{sec:TPchange}.
However, since quench temperature is expected to be sufficiently lower than 5000 K, we start the chemical reaction calculations from 3000 K.
The results from our calculations and \citet{Ishimaru2010-di} both showed that 3000 K is high enough for the gas phase reactions in vapor plumes to reach the equilibrium states (see \Secref{sec:comparison}).
The simulation is terminated when the temperature falls below 300 K, typically lasting for 10 seconds in the case of $r_{\rm imp} = 0.5 \, \rm km$.
%The converted time for $d_{\rm imp} = 1 \, \rm km (??)$ is 10 seconds. Typical chemical reaction steps in our Monte Carlo simulation for one reaction sequence is ?? steps.

\subsection{Evaluation of mole fractions} \label{sec:evolution}

\begin{figure}[h]
\centering
\includegraphics[width=0.9\linewidth]{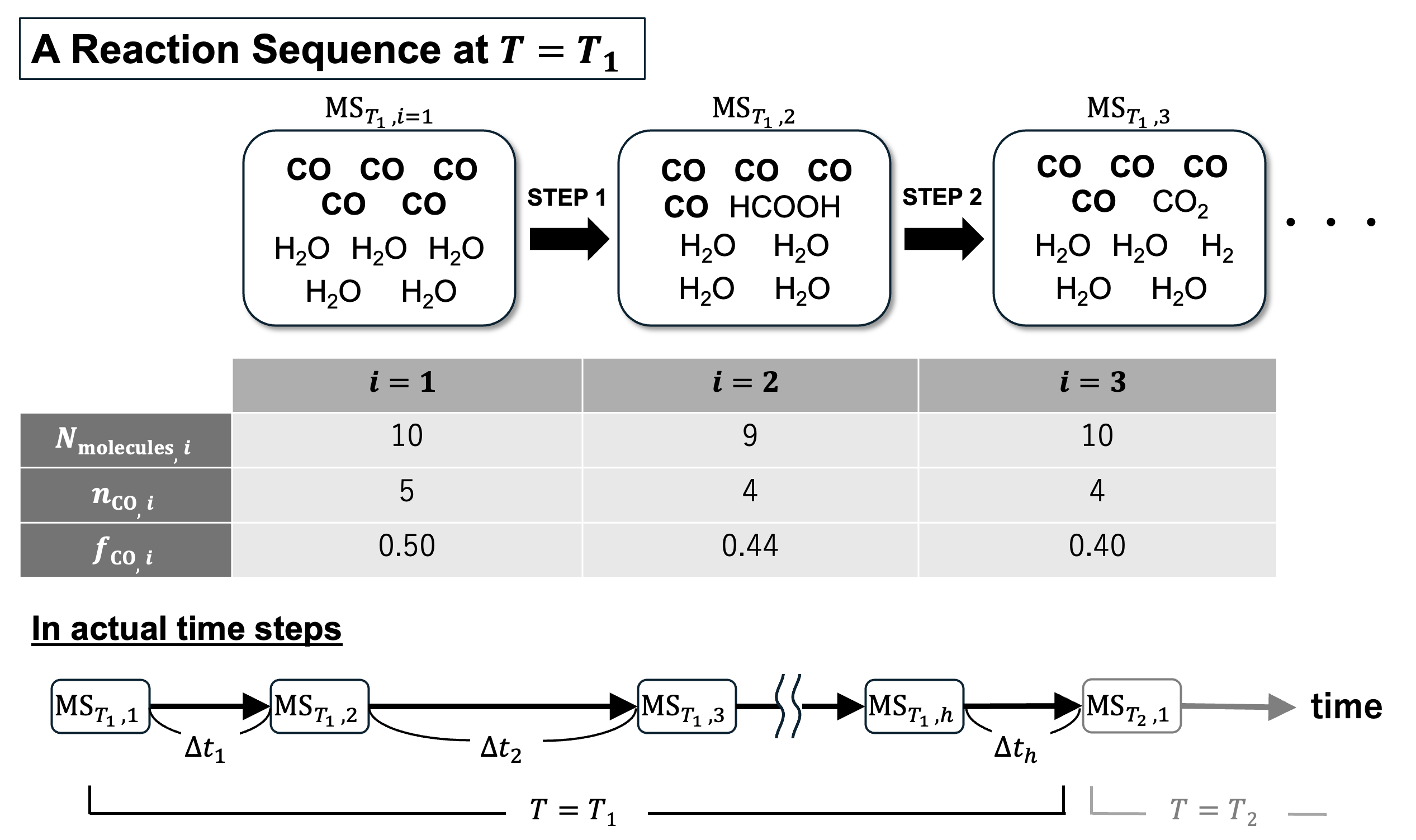}
\caption{
Schematic diagram of a reaction sequence at $T=T_1$.
In this figure, $\rm MS$, $N_{\rm molecules,i}$, $n_{\rm CO,i}$, $f_{\rm CO,i}$ represent "Molecular Set'', the number of molecules in the molecular set, and the number of CO molecules and the mole fraction of CO, respectively.
}
\label{fig:reaction_sequence}
\end{figure}

Here, we describe the method to evaluate mole fractions based on molecular sets and reaction sequences obtained in this chemical reaction simulation.
%Focusing on the $i$-th reaction step, l
Let $N_{\rm molecules,i}$ be the total number of molecules in the 
molecular set before the $i$-th reaction happens, and let $n_{{\rm A} ,i}$ be the number of molecules of a chemical species A in the same set. The mole fraction of A at this step is given by $f_{{\rm A} ,i}=n_{{\rm A} ,i}/N_{\rm{molecules},i}$ as shown in \Figref{fig:reaction_sequence}.

However, in this study, for computational reasons, we calculate the mole fraction representative of each temperature rather than for each individual step. Since the calculations at each temperature consist of multiple steps, the representative mole fraction for a given temperature is taken as a weighted average, accounting for the differences in reaction time $\Delta t_i$ across reaction steps. 

At $T=T_1$, this weighted average of the mole fractions, $F_{\rm A} (T_1)$, is expressed as:
\begin{equation}
    F_{\rm A} (T_1) = \frac{\sum_{i=1}^{n} (\Delta t_i f_{{\rm A} ,i})}{\sum_{i=1}^{n} (\Delta t_i)},
\label{eq:molefrac}
\end{equation}
where $f_{{\rm A} ,i}$ is the mole fraction of species A at the $i$-th step, $n$ represents the total number of reaction steps at $T=T_1$.
As described in \Secref{sec:coupling}, the temperature is reduced by 5 \% with each update, starting at 3000 K in this study. 
At each temperature, the mole fraction of the species of interest is calculated using Eq.~(\ref{eq:molefrac}). 
Since this Monte Carlo simulation consists of multiple independent trials generated with different random number seeds, as mentioned in \Secref{sec:montecarlo}, the mole fractions at each temperature are finally averaged across the individual reaction sequences obtained from these trials.

\section{Results}\label{Result}

\subsection{Time evolution of mole fractions of major species}

First, we focus on the time evolution of mole fractions of the typical major species--$\rm H_2, H_2 O, CO_2, CO, CH_4, N_2$, and $\rm NH_3$--to illustrate the general trends in composition evolution and to confirm the consistency of our results with the results obtained by the previous work \citep{Ishimaru2010-di}. 
%We discuss general trends of time evolution of the major species.

\subsubsection{Comparison with a previous study on cometary impact}\label{sec:comparison}

\begin{figure}
\centering
\includegraphics[width=0.4\linewidth]{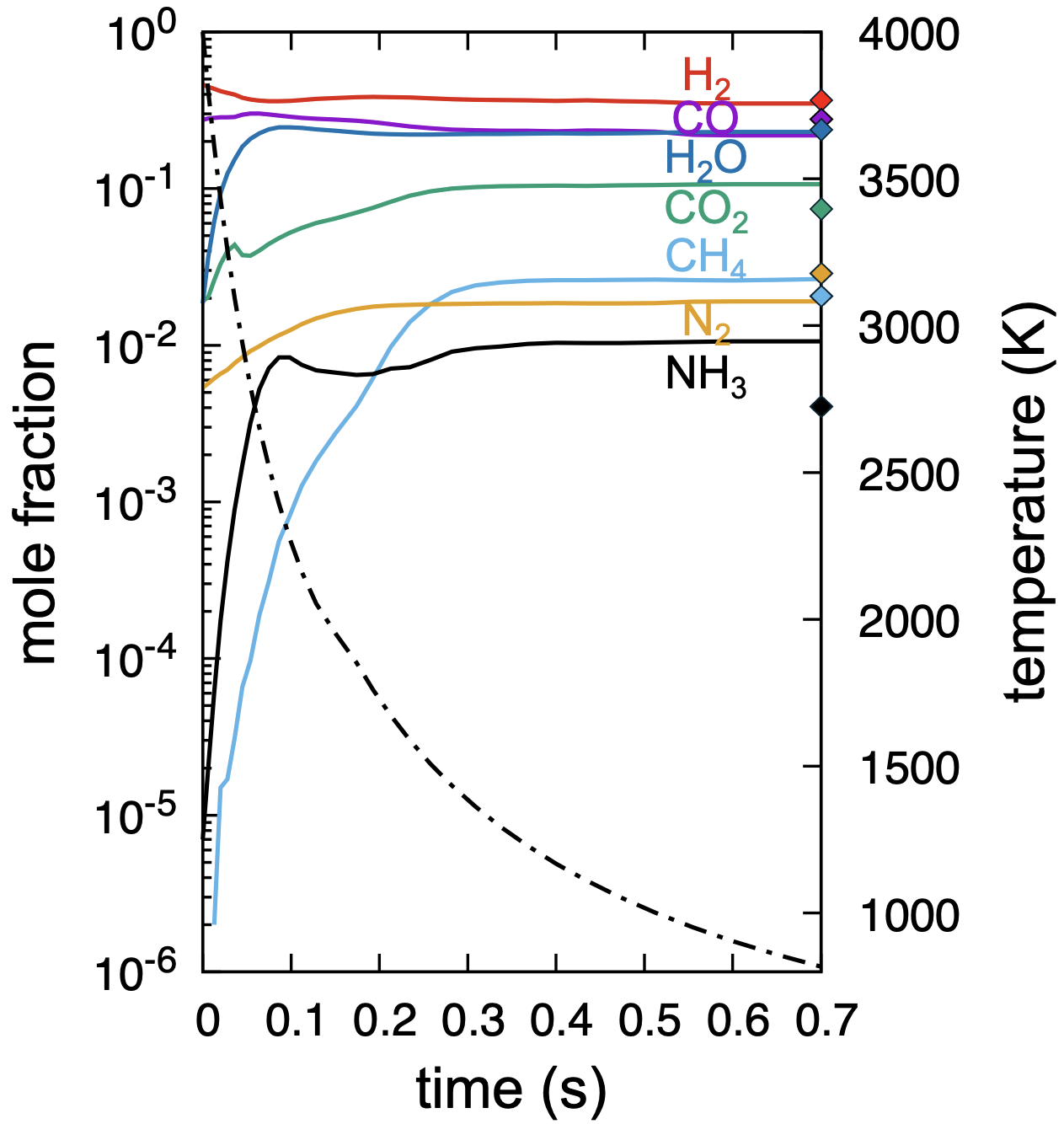}
\caption{Time evolution of mole fractions of major species ($\rm H_2, H_2 O, CO_2, CO, CH_4, N_2$, and $\rm NH_3$; solid lines, left axis) and temperature (dot-dashed line, right axis) for a cometary impact.
Diamonds on the right axis represent the mole fractions of each species in the quenched composition obtained in the previous study \citep{Ishimaru2010-di} (thier Fig.~6).
The atomic composition used is shown in Table~\ref{table:composition} (comet type).
The initial temperature and initial pressure are 4000 K and 30 GPa, respectively, corresponding to the conditions in \citet{Ishimaru2010-di}.
The value of $\beta$ in Eq.~(\ref{eq:LFER}) is set to 300 kJ/mol.
}
\label{fig:Ishimaru}
\end{figure} 

Although comparing our results with experimental studies would be a valuable way to validate the simulations, experimental data are often not sufficiently quantitative or comprehensive to justify theoretical models.
Therefore, in this subsection, we compare our simulation results with those obtained using a different theoretical approach--a kinetic model--which provides a useful benchmark for major molecular species.

\citet{Ishimaru2010-di} simulated impact-induced chemical reactions using a kinetic model (GRI-Mech version 3.0) based on a predefined reaction network, but only for a cometary impact case.
While the present study focuses on asteroidal (rocky-body) impacts, kinetic model simulations for such cases have not yet been published.
Accordingly, we adopt a cometary composition in this subsection to enable direct comparison with the results of \citet{Ishimaru2010-di}.

While the validity of GRI-Mech 3.0 for combustion processes has been demonstrated through comparisons with experimental data, our results, as shown later, demonstrate the synthesis of minor products not included in GRI-Mech 3.0 that could play an important role in amino acid synthesis. 
This is because, unlike kinetic models, our Monte Carlo simulation considers all possible reactions, enabling the synthesis of chemical species excluded from the kinetic model. 
It should be noted that, species with low abundance and/or small impact on the reaction network of the targeted chemical process are usually omitted from kinetic models.

Nevertheless, since such minor chemical species have little effect on the major composition, our results for the evolution of the major species should be consistent with those of \citet{Ishimaru2010-di}.
The prediction of the mole fraction evolution for the major species of the cometary composition (Table~\ref{table:composition}) with our Monte Carlo method is shown in \Figref{fig:Ishimaru}.
The quenched composition of the seven major products is consistent with \citet{Ishimaru2010-di} (their Fig. 6), although some deviations are observed in the early evolution of \ce{H2O}, \ce{CH4}, and \ce{NH3} at temperatures around 4000–2000 K. 
These deviations are likely due to approximations in the Gibbs energy (approximations in enthalpy and entropy changes), but do not significantly affect the quench process, which occurs at lower temperature ($\sim 1000$ K).
%although the result shows that the details of mole fraction evolution of each chemical species deviate from those in \citet{Ishimaru2010-di} (their Fig. 6) by up to an order of magnitude. The deviations are likely due to approximations in the Gibbs energy (approximations in enthalpy and entropy changes, as well as uncertainties in the $\alpha$ and $\beta$ parameters) and differences between the Eyring equation and empirically obtained reaction rate constants.
Thus, Figure~\ref{fig:Ishimaru} supports the validity of our method in capturing overall compositional trends, rather than accurately predicting the exact abundances of individual species.
We also note that, due to the limited information provided in \citet{Ishimaru2010-di}, the $T$-$P$ path used in our calculation for \Figref{fig:Ishimaru} may not fully replicate theirs, potentially contributing to the observed deviations.

While this calculation focuses on a cometary impact, comparisons with kinetic model calculations for chondritic compositions, which are the assumed impactor compositions in this study, are also essential. Preliminary results from such calculations (Miyawaki and Genda, in preparation) demonstrate consistency with our Monte Carlo simulations across CI, LL, and EL compositions. 
Building on the agreement between our Monte Carlo method and the kinetic model for these results for major species mole fractions, we extend this approach to investigate the major species generated by a rocky body impact.
%We emphasize that, as will be shown in Section~\ref{sec:amionacid}, our Monte Carlo method can simulate the production of any minor species, including amino acids, sugars, and their precursors, as it does not exclude any potential reaction pathways.

\subsubsection{Results for CI, LL, and EL chondritic compositions}

\begin{figure}[h]
\centering
\includegraphics[width=1\linewidth]{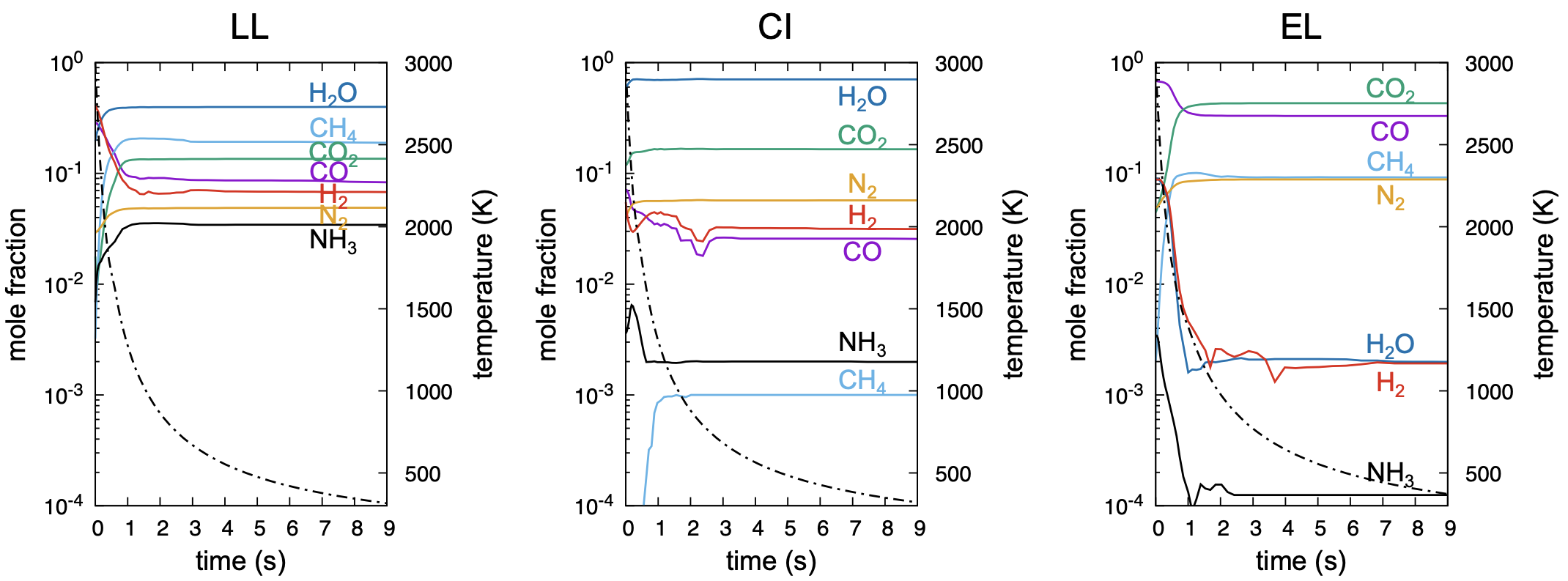}
\caption{Time evolution of mole fractions of major species, $\rm H_2, H_2 O, CO_2, CO, CH_4, N_2$, and $\rm NH_3$ (solid lines) and temperature (dot-dashed line), obtained from the CI, LL, and EL chondritic compositions.
The initial condition of the impact vapor plume is $T_0=5000$ K and $P_0=140$ GPa. Note that the calculations are started from 3000 K ($t=0$) as explained in \Secref{sec:initialcondition}. 
The value of $\beta$ in Eq.~(\ref{eq:LFER}) is set to 300 kJ/mol.
}
\label{fig:molefrac}
\end{figure}

Figure~\ref{fig:molefrac} shows post-impact time evolution of mole fractions of the major chemical species obtained from the CI, LL, and EL chondritic compositions.
Regarding the C/H and O/H ratios of each impactor composition shown in Table~\ref{table:composition}, CI ($\rm C/H = 0.14, O/H = 0.74$) and LL ($\rm C/H = 0.24, O/H = 0.40$) are similar, with CI being slightly oxidizing. 
In contrast, the EL composition ($\rm C/H = 1.83, O/H = 2.08$) differs significantly, being H-poor and extremely C-rich. 
We note that, as these compositions are the equilibrium compositions of degassed volatiles at 1500 K and 100 bar, they represent different redox states from the bulk compositions.
%and do not  the distances of their formation sites.

%The major gases of the CI composition are $\rm H_2 O$ and $\rm CO_2$ throughout the reaction because of its oxidative condition. The major gases of the LL composition are $\rm H_2$ and CO from 3000 K to about 2500 K. After that, they turn into $\rm H_2 O, CH_4$ and $\rm CO_2$ and remain until quenching. The major gases of the EL composition are CO and $\rm CO_2$ due to its carbon-rich composition.

Figure~\ref{fig:molefrac} shows the composition changes cease almost entirely below around 1000 K due to quenching. 
Both the LL and CI compositions result in H$_2$O-rich gas, but in the LL-type, CH$_4$ and CO$_2$ become the next most abundant species, except for the high-temperature region ($3000-2500$ K), where H$_2$ and CO are dominant.
On the other hand, in the more oxidizing CI composition, CO$_2$ becomes the second most abundant species, while the fraction of CH$_4$ is about two orders of magnitude lower than that in the LL-type.

In the EL composition, CO$_2$ and CO become the dominant components after the initial high-temperature region ($3000-2500$ K), where CO, H$_2$, and H$_2$O are prevalent. Despite its very high O/H ratio, CH$_4$ is also relatively abundant ($F_{\rm CH_4} \sim 0.1)$. These characteristics of the EL-type are attributed to its C-rich composition.

For N-bearing species, the mole fraction of $\rm N_2$ is similar across the three compositions ($F_{\rm N_2} \sim 0.1-0.01$) and is always the dominant N-species in their quenched compositions.
In contrast, the mole fraction of $\rm NH_3$ decreases significantly (by up to two orders of magnitude) in the order of LL, CI, and EL compositions, corresponding to the O/H ratio trend.
As will be discussed in \Secref{sec:bio_mechanism}, NH$_3$ is an essential molecule for many biomolecular synthesis pathways. The strong dependence of its mole fraction on impactor composition can suggest the importance of NH$_3$ production as a key process in biomolecular synthesis within impact vapor plumes.

%and from the transition of the major gases in the LL chondritic composition and the CI chondritic composition.
%While some radical species (H, O, OH and HCO radicals) were also produced in all the types with low fraction, they significantly decreased with the drop in temperature.

%\Figref{fig:} also indicates that the quench temperatures depend on the chemical species, varying from ?? K to ?? K. The result that quench temperatures are different among species is consistent with the results \citet{Ishimaru2010-di}. On the other hand, some species exhibited quench temperatures much lower than those estimated in previous studies \citep[e.g.,][]{Gerasimov1999-mh, Fegley1986-vs}. We discuss the quench process and the validity of the quench temperatures shown here in \Secref{sec:quench}.

\subsubsection{Dependence on the model parameter $\beta$} \label{sec:beta}

\begin{figure}[h]
\centering
\includegraphics[width=0.9\linewidth]{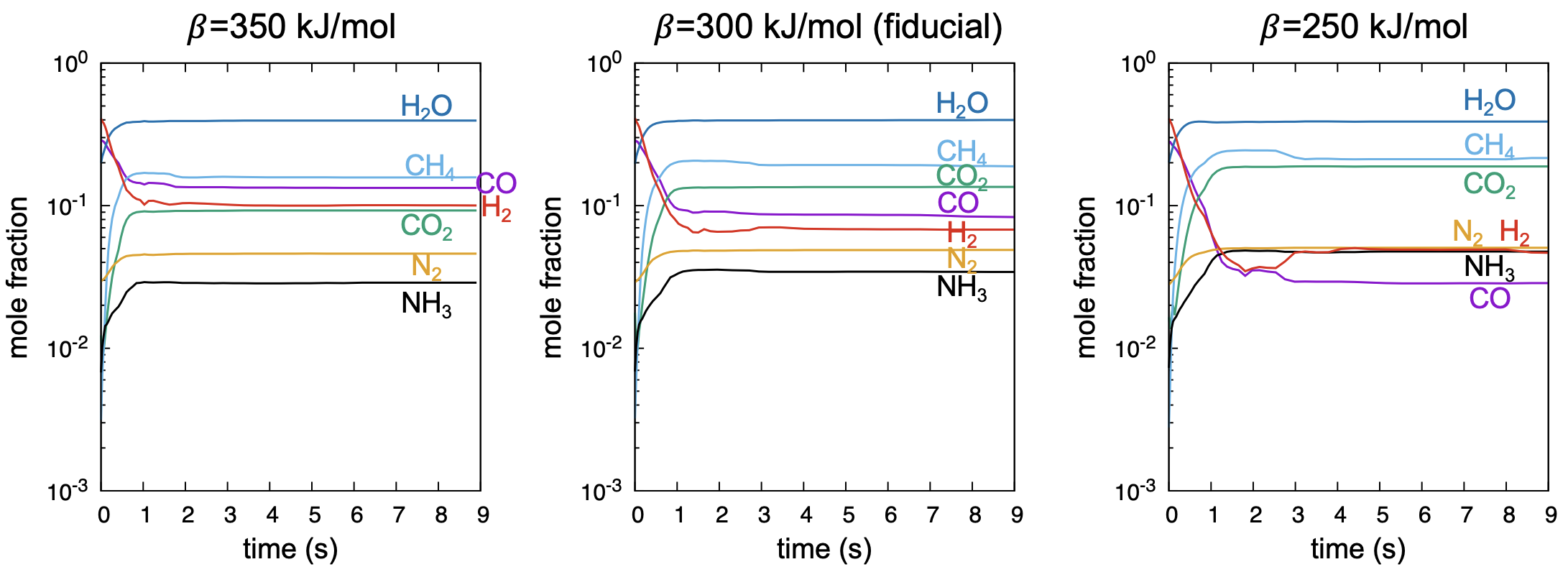}
\caption{Time evolution of mole fractions of major species obtained from the LL-type composition using $\beta=$250, 300, and 350 kJ/mol.
The value of $\beta$ influences the estimation of the standard Gibbs energy of activation (Eq.~(\ref{eq:LFER})), and consequently, the overall reaction rates (Eq.~(\ref{eq:weight})) in this study.
All other calculation settings are identical to those used in \Figref{fig:molefrac}.
}
\label{fig:beta}
\end{figure} 

Figure~\ref{fig:beta} shows the change in mole fractions of major species in the LL composition, calculated with different values of $\beta$, the parameter to evaluate the Gibbs energy of activation in Eq.~(\ref{eq:LFER}).
As Eq.~(\ref{eq:weight}) indicates, the larger the $\beta$ is, the smaller the estimated reaction rate constants are (i.e., the slower the reactions are).
Therefore, the choice of $\beta$ influences the overall reaction rates in the simulations.

On the other hand, equilibrium composition is determined by $\Delta_{\rm r} G$ and is independent of $\beta$ (see Appendix A).
Thus, in the high-temperature region where the equilibrium is reached, the compositions remain the same regardless of the $\beta$ value, as shown in the very early evolution ($\sim 3000-1500$ K) in \Figref{fig:beta}.

Since the cooling timescale is also independent of $\beta$ (Eq.~\ref{eq:t_cool}), increasing $\beta$ slows down the reaction rates without changing how quickly the system cools.
As a result, reactions are quenched earlier (at higher temperatures) when $\beta$ is large, because the slower reactions can no longer keep up with the cooling.
In contrast, when $\beta$ is smaller, reactions proceed more quickly and continue to evolve even at lower temperatures before being quenched.
This allows the composition to shift further, following equilibrium trends associated with decreasing temperature (e.g., \ce{2 CO + 2 H2 <=> CH4 + CO2}).

Nevertheless, within the range of $\beta$ values assumed in this study ($\beta = 250$–$350\ \rm kJ/mol$), these effects do not substantially alter the major components of the quenched composition.
Therefore, the influence on the abundance of other products, including organic compounds discussed in the following sections, is also considered minimal.
Although we focus only on the dependence on $\beta$ in this study, changing the value of $\alpha$ would have a similar effect, as it influences only the reaction rates and quench temperature but not the equilibrium composition.

\subsection{Biomolecules synthesis in impact vapor plumes}\label{sec:amionacid}

In \citet{Ochiai2024}, the previous version of our previous Monte Carlo simulation showed that amino acids and sugars are synthesized by UV irradiation under cryogenic conditions (with $T \sim 50-100 \, \rm K$), corresponding to the surface of icy grains in a protoplanetary disk. 
We also investigated how the synthesis efficiencies of amino acids and sugars depend on the initial atomic ratios of the ice composition, specifically the C/H and O/H ratios.

In contrast, the present simulations do not yield significant mole fractions of amino acids and sugars.
This is because these compounds are thermodynamically unstable at high temperatures and decompose rapidly, even if they are episodically formed.
However, as shown in the following sections, our results show that a sufficient variety of precursors necessary for the formation of biomoecules, such as amino acids, sugars, and nucleobases, are synthesized and remained in the quenched compositions.
Notably, some of these precursors are absent in the previous reaction network used by \citet{Ishimaru2010-di}, meaning that they cannot be produced within that framework. 
We first present the results on the synthesis of organic molecules, including biomolecules' precursors (Secttion~\ref{sec:organics}) and then discuss how biomolecules are synthesized in impact vapor plumes (Section~\ref{sec:bio_mechanism}).
%We first present the results on the precursor synthesis (Secttion~\ref{sec:organics}) and then discuss how amino acids and sugars are synthesized in impact vapor plumes (Section~\ref{sec:aminoacid}).

\subsubsection{Synthesis of organic molecules} \label{sec:organics}

\begin{figure}[h]
\centering
\includegraphics[width=1\linewidth]{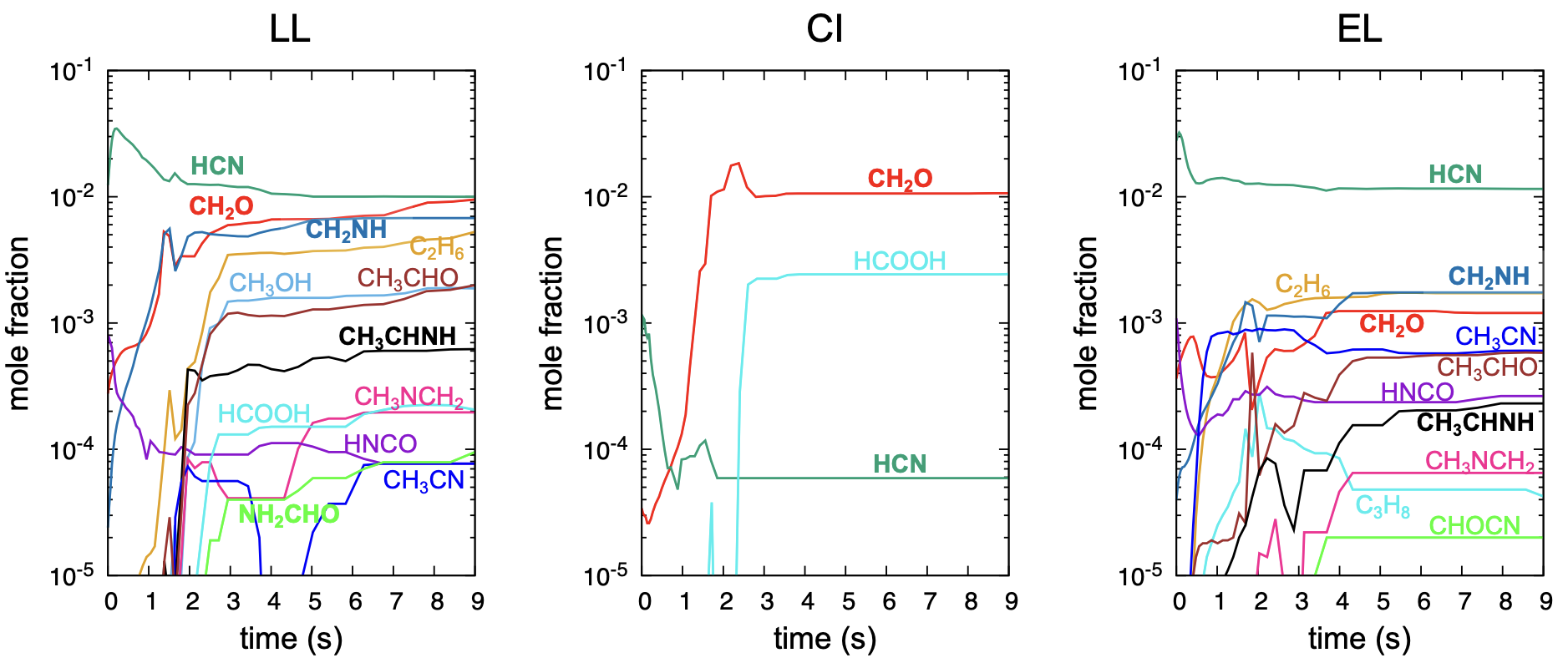}
\caption{Time evolution of mole fractions of organic molecules.
The molecules represented by each chemical formula are as follows: hydrogen cyanide (HCN), formaldehyde (\ce{CH2O}), methanol (\ce{CH3OH}), formic acid (HCOOH), isocyanic acid (HNCO), methanimine (\ce{CH2NH}), ethanimine (\ce{CH3CHNH}), N-methylmethanimine (\ce{CH3NCH2}), acetaldehyde (\ce{CH3CHO}), acetonitrile (\ce{CH3CN}), formamide (\ce{NH2CHO}), formyl cyanide (\ce{CHOCN}), ethane (\ce{C2H6}), propane (\ce{C3H8}).
Molecules considered particularly important for biomolecule synthesis (see \Secref{sec:bio_mechanism}) are indicated with bold lines.
}
\label{fig:organic}
\end{figure} 

Figure~\ref{fig:organic} shows the time evolution of mole fractions of organic molecules produced in the post-impact reactions of the LL, CI, and EL compositions.
We note that the species plotted are not necessarily all organic products but those found to be relatively abundant among about 30 relevant organic molecules examined.
Although isocyanic acid (HNCO) is classified as inorganic compounds, they are also plotted in \Figref{fig:organic} because of their structural similarities to organic molecules.

%While the major organic species in the final composition of the LL chondritic composition are hydeogen cyanide (HCN) and formalydehyde (\ce{CH2O}), methanimine (\ce{CH2NH}), ethane (C$_2$H$_6$), and ethanimine (\ce{CH3CHNH}) are also abundant.Acetaldehyde (C$_2$H$_4$O), propane (C$_3$H$_8$), and methylamine (CH$_3$NH$_2$) are synthesized with lower concentrations as well.In the CI chondritic composition, major organic species were methane (CH$_4$) and hydeogen cyanide (HCN)\:(\Figref{fig:molefrac}).Isocyanic acid, cyanic acid, formamide (HCONH$_2$), and formic acid (HCOOH) are also relatively abundant, reflecting its oxidative composition.The EL chondritic composition shows the most diverse organic products among these three composition types.The major species are methane, hydrogen cyanide, formalydehyde (\Figref{fig:molefrac}), ethane, methanimine, acetonitrile (CH$_3$CN), ethanimine, propane, and isocyanic acid, exhibiting the products rich in formyl, cyano, amino, and imine groups.Cyanic acid, acetaldehyde, cyanamide, and cyanoacetylene are found only at high temperatures.

In the LL chondritic composition, a variety of organic molecules are produced.
The major species amoung them are hydrogen cyanide (HCN), formalydehyde (\ce{CH2O}), methanimine (\ce{CH2NH}), and ethane (\ce{C2H6}).
Other products include methanol (\ce{CH3OH}), acetaldehyde (\ce{CH3CHO}), ethanimine (\ce{CH3CHNH}), N-methylmethanimine (\ce{CH3NCH2}), formic acid (HCOOH), formamide (\ce{NH2CHO}), acetonitrile (\ce{CH3CN}), and isocyanic acid (HNCO). 
As these products suggest, nitrogen-containing organic compounds are produced more diversely and abundantly in the LL type compared to the other two types, enabling the synthesis of species like formamide, which is absent in the other two compositions.

The EL chondritic composition generates organic compounds that are somewhat similar to those produced by the LL composition.
The most abundant species is HCN, and the fractions of all other products--methane (\ce{C2H6}), methanimine (\ce{CH2NH}), formalydehyde (\ce{CH2O}), acetonitrile (\ce{CH3CN}), acetaldehyde (\ce{CH3CHO}), isocyanic acid (HNCO), ethanimine (\ce{CH3CHNH}), N-methylmethanimine (\ce{CH3NCH2}), propane (\ce{C3H8}), and formyl cyanide (\ce{CHOCN})--are approximately one order of magnitude, or even more, lower than that of HCN.
Despite its high O/H ratio, the carbon-rich composition of the EL type promotes organic synthesis, resulting in the relatively abundant formation of a C3 compound (\ce{C3H8}).
Moreover, while the EL type exhibits a significantly low abundance of NH$_3$ among the three compositions (\Figref{fig:molefrac}), the efficient production of HCN presumably promotes the synthesis of N-containing organic compounds, including acetonitrile and formyl cyanide.
%the competition between C-rich (advantage for organic production) and shortage of \ce{NH3} (nearly N source available for organic production)

In the CI composition, only hydrogen cyanide (HCN), formaldehyde (\ce{CH2O}), and formic acid (HCOOH) are observed with significant fraction, indicating the most limited organic compound production among the three compositions. 
Compared to the LL type, which has a similar atomic ratio, the CI composition is more C-poor and oxidizing, resulting in a significantly lower production rate of reactive species such as NH$_3$ and CH$_4$. 
This is likely to inhibit the overall synthesis of organic compounds in the CI composition.

%In all the three types, C1 compounds such as formaldehyde and hydrogen cyanide are the most abundant organic molecules due to thermodynamical stability at high temperatures.Nitrogen organic compounds are also produced in all the types, such as hydrogen cyanide, methanimine and acetonitrile.On the other hand, different kinds of organic molecules are identified depending on the composition types, such as methylamine in LL-type, formic acid in CI-type, and cyanoacetylene in EL-type.
It is worth noting that we confirmed a decrease in the O/H ratio of the vapor composition leads to an increase in the mole fraction of HCN for each impactor material.
However, it is also evident that O/H is not the sole factor controlling HCN production, as demonstrated by the fact that the EL composition yields a higher HCN mole fraction than the CI composition, despite having a higher O/H ratio.
Other parameters, such as C/N and/or C/O ratios, may also play important roles in regulating HCN formation.
Understanding these dependencies requires further study, which is left for future work.

While, in these simulations, we detect a variety of complex organic molecules, including amino acids, particularly at high temperatures ($> 1500$ K), their concentrations are extremely low. 
Such complex organic molecules, consisting of a large number of atoms, are thermodynamically unstable and rapidly decompose even if they happen to be synthesized. 
Many complex organic molecules with extremely short lifetime (on the order of less than 1 nanosecond) have been also observed in AIMD simulations of post-shock chemical reactions at temperatures about 3000–4000 K \citep{Goldman2010-np}.
%Consequently, as our results show, it is highly unlikely that complex organics  synthesized in gas-phase reactions remain in detectable amounts in the quenched composition.

We stress that most of the organic compounds identified in this study (\Figref{fig:organic}) are not included in the reaction network used in the previous study \citep{Ishimaru2010-di}, and some of them can play crucial roles in biomolecules synthesis.
For example, methanimine (CH$_2$NH), produced in the LL and EL compositions, has been proposed as a precursor to glycine (the simplest amino acid) via the Strecker reaction, one of the plausible reaction mechanisms for prebiotic amino acids synthesis.
Formamide (\ce{NH2CHO}) has also been suggested as a key molecule for the prebiotic synthesis of nucleobases \citep[e.g., ][]{Saladino2001-un}.
We discuss potential mechanisms for biomolecules synthesis in more detail in \Secref{sec:bio_mechanism}.

Since GRI-Mech, the chemical kinetic model used in \citet{Ishimaru2010-di}, was primarily designed for modeling combustion processes, it selectively focuses on species and pathways relevant to combustion reactions. 
As a result, most kinds of trace organic compounds shown in \Figref{fig:organic} are not considered in the reaction network of GRI-Mech. 
Accordingly, \citet{Ishimaru2010-di} were not able to predict the formation of these compounds.
Our results thus provide the first theoretical prediction of diverse organic compound production and the synthesis of precursors to biomolecules in impact vapor plumes that produced by various chondritic materials.
This underscores the significance of the Monte Carlo simulation approach, which does not rely on predefined chemical species and reaction pathways, enabling the inclusion of all potential products.

\subsubsection{Possible mechanisms for the syntheses of amino acids, sugars, and nucleobases} \label{sec:bio_mechanism}

As mentioned above, the present calculations show that no biomolecules remain in the quenched compositions of impact vapor plumes. 
However, our results show that several molecules, which serve as starting materials or intermediates in biomolecule synthesis pathways, are abundantly produced.
As the temperature of the impact vapor plume decreases, the condensation of H$_2$O and the dissolution of the reactants of those reactions may allow biomolecule synthesis to proceed in the aqueous solution. 
While our current model, which focuses on gas-phase chemical reactions, does not account for condensation processes and liquid-phase reactions, we assume the formation of aqueous solutions after quenching and outline the potential biomolecule synthesis pathways based on the products identified in this study.

One such pathway is the Strecker reaction, a potential prebiotic synthesis route for $\alpha$-amino acids \citep[e.g., ][]{Magrino2021-og}. 
This reaction involves two main stages: (1) formation of $\alpha$-amino nitrile from the starting molecules (aldehydes, NH$_3$, and HCN), and (2) hydolysis of $\alpha$-amino nitrile to $\alpha$-amino acid, as follows:
\begin{align}
& \ce{R-CHO + NH3 + HCN ->[-H2O] R-CH=NH}\rm{\ (imine)} \ce{+HCN -> R-CH(NH_2)CN}  \rm{\ (\alpha-amino nitrile)} \label{eq:Strecker1}\\ 
& \ce{R-CH(NH_2)CN ->[+2 H2O][-NH3] R-CH(NH_2)COOH} \rm{\ (\alpha-amino\ acid)}.
\label{eq:Strecker2}
\end{align}
Our results indicate that the initial molecules for this reaction--anmonia (NH$_3$), hydeogen cyanide (HCN), and formaldehyde (CH$_2$O)--are simultaneously produced in all the compositions (Figs.~\ref{fig:molefrac} and \ref{fig:organic}).
Furthermore, our results also show that imine compounds, the intermediates of reaction (\ref{eq:Strecker1}), such as methanimine (\ce{CH2NH}) and ethanimine (\ce{CH3CHNH}) are produced in impact vapor plumes (\Figref{fig:organic}), possibly advancing amino acid synthesis. 
Although \cite{Magrino2021-og} explored the free energy landscape of each step in the Strecker reaction, including reactions involving imines, the current understanding of how these reactions proceed abiotically remains limited.
To clarify the role of imine intermediates in prebiotic amino acid synthesis, further kinetic data and experimental studies under plausible prebiotic conditions will be essential.

While the Strecker reaction can only synthesize $\alpha$-amino acids, $\beta$-amino acids have also been synthesized in shock-recovery experiments \citep{Furukawa2008-hx, Furukawa2015-xx, Takeuchi2020-yr}.
The ammonia-involved formose-type reaction, initiated from ammonia and formaldehyde in alkaline conditions, has been proposed as a synthesis pathway for amino acids, including $\beta$-amino acids. 
Our results indicate that both of these starting molecules (CH$_2$O and NH$_3$) are synthesized in impact vapor plumes (Figs.~\ref{fig:molefrac} and \ref{fig:organic}), suggesting that amino acids could also be produced through the formose-type reaction. 
Similarly, such conditions may enable the progression of the formose reaction, which synthesizes sugars through the condensation of formaldehyde \citep[e.g., ][]{Breslow1959-lf, Bris2024-qh}, implying the potential synthesis of sugars as well.

In the prebiotic synthesis of nucleobases, HCN is a critical starting material. 
Multiple pathways for nucleobase formation from HCN have been proposed \citep[e.g., ][and references therein]{Kitadai2018-nc}, including HCN polymerization and reactions with NH$_3$. 
Both HCN and NH$_3$ are detected in all the three compositions, suggesting the potential for nucleobase synthesis in impact vapor plumes.
However, the abundances of HCN and NH$_3$ vary significantly among the compositions: the EL composition is most favorable for HCN, while the LL composition favors NH$_3$ production.
These differences are expected to lead to substantial variations in nucleobase production rates.
Additionally, formamide synthesized in the LL composition has been proposed as another important precursor for nucleobase synthesis \citep[e.g., ][]{Saladino2001-un, Roy2007-yx}, potentially further facilitating their formation. 
Formamide is detected exclusively in the LL composition, the most reducing among the three and both HCN and NH$_3$ are found to be abundant in the LL composition compared to the other two. 
These findings may align with previous experimental results that showed higher nucleobase production in reducing impactor compositions \citep{Furukawa2015-xx}.
Further quantitative evaluations of the production rates of these chemical species and parameter surveys using a broader range of compositions will be necessary.

%HCN and formamide have also been identified as important precursor molecules for nucleobase synthesis \citep[e.g., ][]{Oro1961-lh, Saladino2001-un, Roy2007-yx}. Although HCN is produced in all the three types of compositions (\Figref{fig:organic}), other factors that may influence the production rate of nucleobases, such as the presence of formamide and/or the concentration of NH$_3$, differ significantly among them.
%Our results show that formamide is \brown{the} most abundantly synthesized in the oxidizing CI composition. If formamide is a primary starting material for nucleobase synthesis, this would be inconsistent with the results of \citet{Furukawa2015-xx}, which showed that nucleobases were more abundantly synthesized in reduced compositions. Therefore, based on our current results, it is expected that HCN plays a more significant role in nucleobase synthesis. Further quantitative evaluations of the production rates of these chemical species and parameter surveys using a broader range of compositions will be necessary.

Finally, it should be emphasized that imine compounds, intermediates in the Strecker reaction, and formamide, a potential precursor to nucleobases, were not included in the reaction networks of previous studies. 
Since our Monte Carlo method does not predefine the reaction network and includes all possible reactions and potential products, our simulation was able to find these compounds.
These molecules may be synthesized in relatively high abundances within impact vapor plumes and are critical for understanding the pathways that lead to biomolecule synthesis in such environments.

\section{Discussion}\label{Discussion}

\subsection{Dependence on impactor size and impact velocity}\label{sec:impactorsize}

\begin{figure}[h]
\centering
\includegraphics[width=0.9\linewidth]{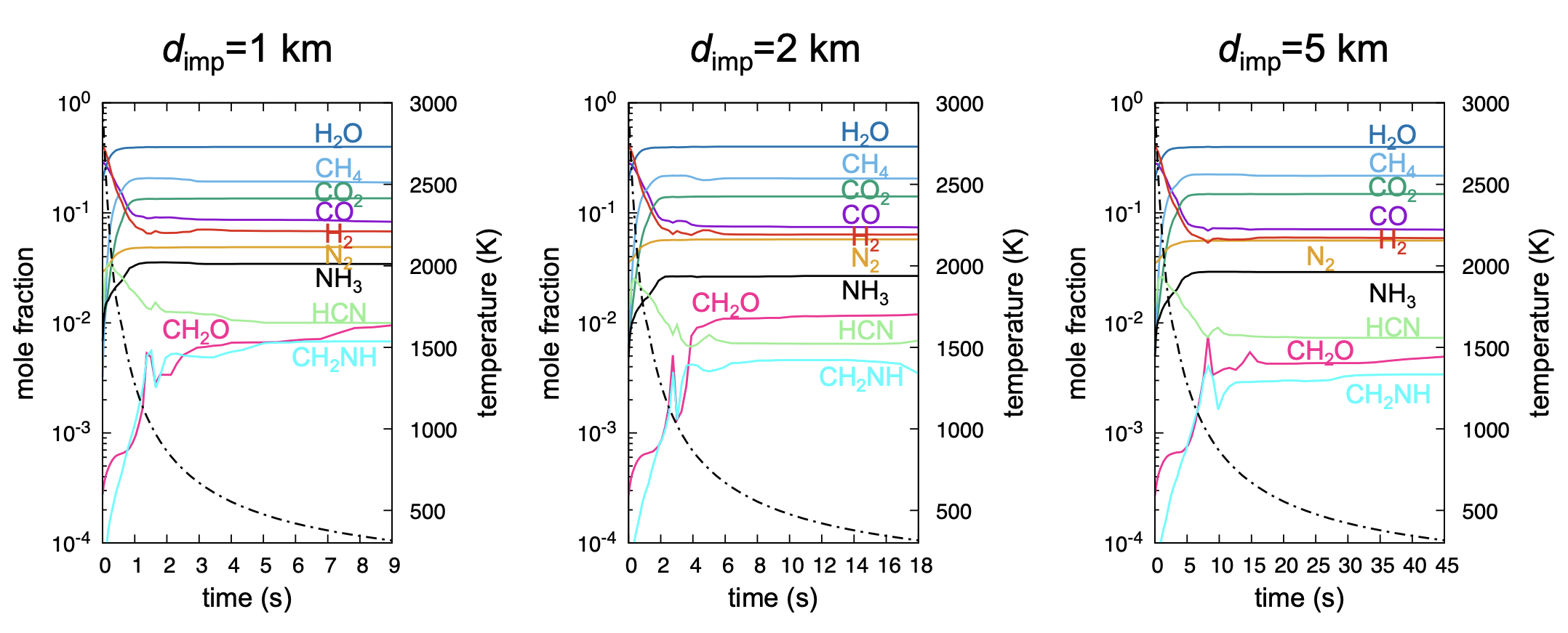}
\caption{Time evolution of mole fractions of $\rm H_2, H_2 O, CO_2, CO, CH_4, N_2$, $\rm NH_3$, HCN, $\rm CH_2O$, $\rm CH_2NH$ (solid lines) and temperature (dot-dashed line), obtained from the LL composition with impactor diameters of 1 km, 2 km, and 5 km.
}
\label{fig:d_imp}
\end{figure} 
\begin{figure}[h]
\centering
\includegraphics[width=1\linewidth]{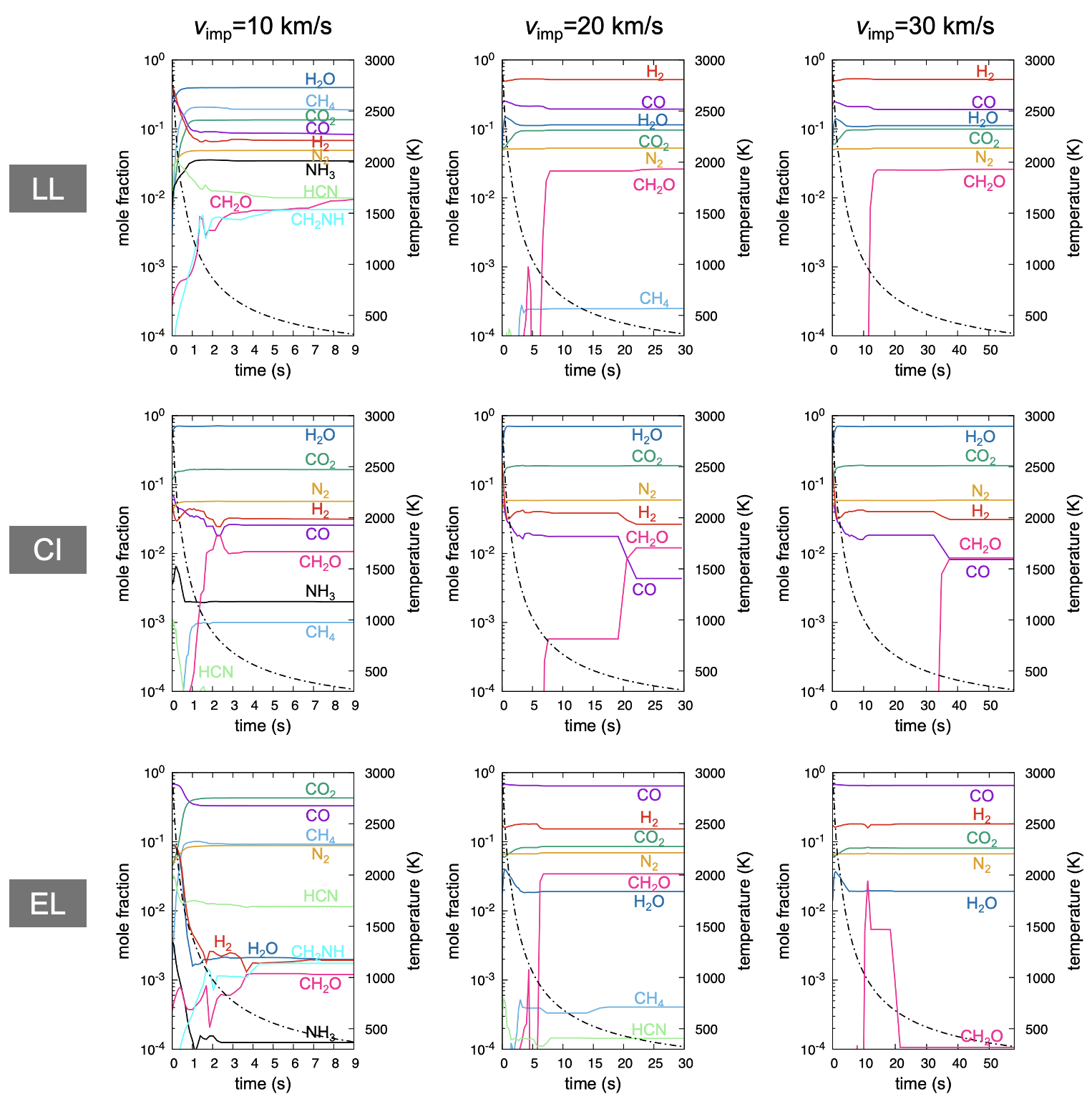}
\caption{Time evolution of mole fractions of $\rm H_2, H_2 O, CO_2, CO, CH_4, N_2$, $\rm NH_3$, HCN, $\rm CH_2O$, $\rm CH_2NH$ (solid lines) and temperature (dot-dashed line), for impact velocities of 10 km/s, 20 km/s, and 30 km/s.
}
\label{fig:v_imp}
\end{figure}

In this study, we assumed an impactor size of 1 km and an impact velocity of 10 km/s as the fiducial case.
Here, to investigate the dependence of the results on these parameters, we focus on the changes in the mole fractions of seven major species ($\rm H_2, H_2 O, CO_2, CO, CH_4, N_2$, and $\rm NH_3$), as well as three molecules important for biomolecule synthesis (hydrogen cyanide (HCN), formaldehyde (\ce{CH2O}), and methanimine (\ce{CH2NH})).
Each simulation starts at 3000 K (see \Secref{sec:initialcondition}), and the value of $\beta$ in Eq.~(\ref{eq:LFER}) is set to 300 kJ/mol.

Impactor size, which influences the initial size of the vapor plume and thus the cooling timescale (\Eqref{eq:t_cool}), does not significantly affect the results within the examined range of $d_{\rm imp}=1-5$ km as shown in \Figref{fig:d_imp}.
These results suggest that changes in the cooling timescale associated with the range of impactor sizes have a negligible effect on the quench temperature.

In contrast, impact velocity, which alters the initial temperature (\Eqref{eq:post_impT}), pressure (\Eqref{eq:post_impP}), and cooling timescale of the plume (\Eqref{eq:t_cool}), can have a substantial effect on the vapor plume composition (\Figref{fig:v_imp}).
Lower impact velocities result in higher pressures at a given temperature, which generally facilitates the formation of larger molecules.
%They also lead to shorter cooling timescales, which, based on our earlier findings, would have a relatively minor effect on the resulting composition.
For the LL and EL compositions, higher impact velocities lead to notable increases in \ce{H2} and CO, accompanied by decreases in \ce{CH4}, HCN, \ce{NH3}, and methanimine.
The mole fraction of \ce{CH2O} increases with impact velocity in the LL case, whereas the clear trend is not found in the EL case.
The CI composition remains largely unchanged across different impact velocities, except for the drastic decease in \ce{CH4}, \ce{NH3}, and HCN.

These results indicate that the dependence on impact velocity varies significantly depending on the initial composition and the molecular species formed.
From the perspective of biomolecule synthesis, lower-velocity impacts are expected to be generally more favorable, as they yield higher molar fractions of \ce{NH3}, HCN, and methanimine.

\subsection{Possible influence of approximations on simulation results}\label{sec:approx}
%Some of the quench temperatures shown in \Figref{fig:molefrac} are much lower than the quench temperatures of 1000 K - 3000 K, that were estimated in previous studies \citep[e.g.,][]{Gerasimov1999-mh, Fegley1986-vs}. They evaluated quench temperature based on the cooling timescale and the reaction rate of decomposition of the chemical species at the temperature.\citet{Fegley1986-vs}
%As explained in \Secref{sec:beta}, quenching temperatures observed in our simulations vary by $\sim 100$ K depending on the model parameter $\beta$ (250-350 kJ/mol in this study).Therefore, we do not focus on the quenching temperatures, but 

In this study, several approximations were made in evaluating reaction rates. 
The dependence of reaction rates on third-body concentrations are assumed to be negligible under the impact vapor plume conditions, as mentioned in \Secref{sec:reactionrates}.
The catalytic effects of solid components will also be discussed in \Secref{sec:rock}.

Equilibrium composition, which is determined by the Gibbs energy of the system, does not depend on individual reaction rates. 
Thus, the uncertainties of our results in high-temperature regions where equilibrium is achieved arise from only approximations for enthalpy change and entropy change.
Among the cases examined, enthalpy changes show maximum discrepancies of several tens of kJ/mol, while entropy differences remain within 20 J/mol·K\footnote{A comparison was made between standard reaction enthalpies and entropies at 300 K calculated using GRI-Mech 3.0 data and those estimated by our model. For example, in the reaction \ce{N2 + 3 H2 -> 2 NH3}, the enthalpy difference is 37 kJ/mol and the entropy difference is 14 J/mol·K. In \ce{CH3OH -> CO + 2 H2}, the enthalpy difference is 29 kJ/mol and the entropy difference is 11 J/mol·K. In \ce{CH2O + H2 -> CH3OH}, the enthalpy difference is 38 kJ/mol and the entropy difference is 6 J/mol·K.}.
Such discrepancies can lead to inaccuracies in the relative mole fractions among products.
However, for the formation of organic molecules in impact vapor plumes--the focus of this study--the influence of such errors in the enthalpy and entropy of individual molecules should be small since our model successfully reproduces the general trends in enthalpy and entropy differences between simple molecules and large molecules, as well as the temperature dependence.

While equilibrium composition is determined solely by the Gibbs energy, the quenching process depends also on reaction kinetics.
In this study, reaction rates are estimated based on the Eyring equation (\Eqref{eq:Eyring1}), and the value of the standard Gibbs energy of activation $\Delta^\ddag G^{\circ}$, which predominantly determines the reaction rates, contains uncertainty due to the parameters $\alpha$ and $\beta$ used in its estimation (\Eqref{eq:LFER}). 
While these parameter uncertainties do not impact the equilibrium composition itself, they influence the reaction rate and, consequently, the timing of quenching, as described in \Secref{sec:beta}.
Furthermore, many of the rate constants used in the kinetic model (GRI-Mech 3.0) employed by \cite{Ishimaru2010-di} deviate from the Eyring equation.
As a result, we found that the empirical rate constants used in GRI-Mech 3.0 sometimes differ significantly from those predicted by our model, depending on temperature.
Nevertheless, the overall consistency between our results and the kinetic model may be attributed to the fact that the equilibrium composition at high temperatures is independent of reaction rates, and that the final quench composition is largely determined by this high-temperature equilibrium state.

We emphasize that this simulation is not intended to accurately predict the production rates of individual molecules.
Importantly, the various organic compounds observed in this study are challenging to assess using conventional methods, highlighting that this approach--while compromising some level of accuracy--has enabled the investigation of such diverse product distributions.

\subsection{Effects of rock-forming elements}\label{sec:rock}
In the current simulations, we ignored rock-forming elements such as Si, Al, Cl, Fe, Mg, Ni and others.
While their effects on the gas-phase composition of impact vapor plumes is partly taken into account when setting the initial conditions (see \Secref{sec:initialcondition}), some rock-forming elements may catalyze specific chemical reactions \citep{Shimamura2019-ll, Shimamura2016-jq}.
Catalysts, which reduce activation energies and accelerate the reactions, should not have any effects to the chemical compositions in high temperature region since the reaction rates are fast enough to reach equilibrium compositions.
However, the presence of catalyst would become important when quenching starts to occur, i.e., the reaction timescale and cooling timescale become comparable.

\citet{Shimamura2019-ll} and \citet{Shimamura2016-jq} investigated shock-induced reactions triggered by meteorite impacts using AIMD simulations.
They suggested that the adsorption of simple carbon and nitrogen species on meteoritic iron efficiently promotes the formation of C-H bonds and NH$_3$, which are important for organic synthesis, via processes similar to the Fischer–Tropsch process.
Experimental studies have also reported the efficient production of CH$_4$ from CO and H$_2$ by Fischer-Tropsch catalysis at around $T=500-600$ K and $P\lesssim 1$ bar \citep{Sekine2005-ov}, which could promote further organic synthesis.
However, a previous study simulating a cometary impact \citep{Martins2013-pz} reported that volatile components alone could produce amino acids, supporting our prediction that amino acids can be synthesized without the catalytic action of metal components.

\section{Conclusion}\label{Conclusion}
 
In this study, we investigated the chemical reactions occurring within the impact vapor plume generated by an asteroid impact on early Earth. 
Focusing on biomolecules synthesized in previous impact simulation experiments \citep[e.g.,][]{Furukawa2008-hx, Furukawa2015-xx, Takeuchi2020-yr}, we aimed to elucidate the organic synthesis processes taking place in the impact vapor plume.

To comprehensively explore the complex chemical pathways involved in organic molecule synthesis, we employed a chemical reaction simulation using Monte Carlo method, developed in our previous work \citep{Ochiai2024}. 
This approach enables the inclusion of all possible reaction pathways without relying on a predefined reaction network by estimating reaction rates based on approximate Gibbs energy changes for each reaction.
Additionally, to account for the rapid changes in temperature and pressure due to the adiabatic expansion of the vapor plume after an impact, we developed a method to couple temperature and pressure changes with chemical reaction calculations.

We performed chemical reaction simulations of impact vapor plumes using three different impactor compositions: LL, CI, and EL types. 
The results showed that a wide variety of organic molecules was synthesized, with the specific types and abundances varying significantly depending on the impactor material (LL, CI, and EL types).
Importantly, these organic products include key precursor molecules essential for the formation of biomolecules such as amino acids, sugars, and nucleobases.
This suggests that asteroid impacts on early Earth may have played a critical role in providing the building blocks of life.

However, throughout the post-shock reactions, the concentrations of biomolecules themselves remained extremely low, suggesting that biomolecule synthesis does not occur directly through gas-phase reactions within impact vapor plumes. 
Instead, our results support a scenario in which precursor molecules synthesized in the gas phase are subsequently concentrated in aqueous solutions following the condensation of H$_2$O gas as the plume cools. 
Within these aqueous environments, these precursors may undergo further reactions, such as the Strecker reaction or the formose-type reaction, ultimately leading to the formation of biomolecules.
Notably, among these precursor molecules, imine compounds and formamide were absent in the reaction network used in \cite{Ishimaru2010-di}, and their synthesis was theoretically predicted here for the first time.
In order to theoretically investigate synthesis of low-abundance complex organic molecules including biomolecules, it is essential to compare kinetic models with non-predefined reaction network approaches, such as our Monte Carlo model, as well as to compare theoretical studies with experiments.    

While various environments, such as the atmosphere and hydrothermal vents, have been proposed as sites for the formation of biomolecules and their precursors, our results show that impacts can also generate a wide range of organic compounds, including key precursors for biomolecule synthesis. Although we do not claim that impact vapor plumes are the most favorable setting for biomolecule synthesis, our results highlight their possible contribution and provide a basis for comparison with other origin-of-life scenarios.

In addition, while we assumed mixing of the impact vapor plumes with a pure nitrogen atmosphere, the effect of atmospheric composition on the organic synthesis in impact vapor plumes is limited under low atmospheric mixing conditions in this study. This implies that our results may be applicable to a variety of planets and moons beyond Earth. 
%In such cases, impacts could have similarly generated precursor molecules elsewhere in the solar system. 
Although further investigation is required, our findings support the possibility that impact-induced synthesis of biomolecular precursors--and potentially biomolecules themselves--may have occurred on other bodies as well.

\section{Acknowledgments}
We would like to thank Akira Miyoshi, Yoshihiro Furukawa, Hidenori Genda, and Seiichi Miyawaki for their insightful discussions and valuable feedback.
This research is supported by JSPS Kakenhi grant 21H04512 and JST SPRING, Japan Grant Number JPMJSP2106 and JPMJSP2180.

\printcredits

%\bibliographystyle{cas-model1-num-names}
\bibliographystyle{cas-model2-names}

\bibliography{reference}

\bio{}

\endbio

\section*{Appendix A. Estimation of the model parameters $\alpha$ and $\beta$}

Here, we consider the reaction \ce{A <=>[1][2] B}.
If we assume that the linear relationship $\Delta^\ddag G^{\circ} = \alpha \Delta_r G^{\circ} + \beta$ applies to each reaction, the standard Gibbs energies of activation for the forward reaction (1) and the reverse reaction (2) can be expressed as follows.
\begin{equation}
    \Delta^\ddag G^{\circ}_1= \alpha_1 \Delta_r G^{\circ}_1 + \beta_1,
\end{equation}
\begin{equation}
    \Delta^\ddag G^{\circ}_2= \alpha_2 \Delta_r G^{\circ}_2 + \beta_2.
\end{equation}

When the reaction rate constants are described by the Eyring equation (\Eqref{eq:Eyring1}), the ratio of the rate constants $k_1$ and $k_2$ can be expressed as: 

\begin{equation}
    \frac{k_1}{k_2}=\frac{\frac{\kappa k_{\rm B} T}{h} \exp\left(-\frac{\Delta^\ddag G^{\circ}_1}{RT}\right)}{\frac{\kappa k_{\rm B} T}{h} \exp\left(-\frac{\Delta^\ddag G^{\circ}_2}{RT}\right)}=\exp\left(-\frac{(\alpha_1+\alpha_2)\Delta_r G^{\circ}_1+(\beta_1-\beta_2)}{RT}\right),
\end{equation}
where $\Delta_r G^{\circ}_1$ is the standard reaction Gibbs energy for the reaction \ce{A -> B}.
This derivation employs the relationships $\Delta_r G^{\circ}_2=-\Delta_r G^{\circ}_1$ and $\Delta_r G^{\circ}=\Delta^\ddag G^{\circ}_1-\Delta^\ddag G^{\circ}_2$.
Based on equilibrium theory and kinetics, this ratio $k_1/k_2$ is equal to $\exp\left(-\frac{\Delta_r G^{\circ}_1}{RT}\right)$.
By comparing these expressions, the following conditions for $\alpha$ and $\beta$ are obtained:
\begin{equation}
    \alpha_1 + \alpha_2=1,
\end{equation}
\begin{equation}
    \beta_1 - \beta_2 =0.
\end{equation}

In this study, we assume the simplest case, where $\alpha_1 = \alpha_2=0.5$.
While the above equation does not constrain the value of $\beta$, we adopt a range of $\beta=250 - 350\ \rm kJ/mol$, based on comparisons between our calculations and the results of \citet{Ishimaru2010-di}.
Additionally, we assume that the values of $\alpha$ and $\beta$ remain constant across all reactions for simplicity.

\section*{Appendix B. Derivation of a time step in the chemical reaction calculation}
\label{appendix}

In our model, at each reaction step, multiple candidates for the next reaction are enumerated.
The set of these candidate reactions varies not only between simulations but also between individual trials or steps, even within the same simulation.
This leads to the possibility that, in one step, all candidate reactions have high reaction rates, while in another step, all candidates exhibit low reaction rates.
Consequently, the time steps corresponding to these steps differ significantly.
For instance, in a system where reaction rates are generally high, such as at high temperatures, the time step for a single reaction step is expected to be much shorter than in a system where reaction rates are low, such as at low temperatures.
However, time step evaluation was not conducted in the previous study by \citet{Ochiai2024}, as their focus was on systems with constant temperature and where reaction rates did not vary significantly from step to step due to the predominance of barrierless reactions. 
In this study, we quantitatively evaluate the time step associated with each reaction step by comparing it with the kinetic model scheme.

In kinetic models, the change in concentration of chemical species $J$ in one time step $\Delta t$, $\Delta \lbrack J \rbrack$, is written as:
\begin{equation}
    \Delta \lbrack J \rbrack= (r_{J+}-r_{J-})\Delta t,
\label{eq:kinetic_dJ}
\end{equation}
where $r_{J+}$ and $r_{J-}$ are rates of the reactions that produce and consume $J$, respectively.

In our Monte Carlo calculation, the change in concentration of chemical species $J$ in the reaction step $i$ of a given reaction sequence (hereafter, subscript $i$ means the value at the $i$-th step) is written as:
\begin{equation}
    \Delta \lbrack J \rbrack_i= \frac{(r_{J+,i}-r_{J-,i})}{\sum r_i} \Delta c_i
    %{\sum_{m=1}^{n_{\rm{cand},i}}r_m} 
\label{eq:Monte_dJ}
\end{equation}
where $\sum r_i$ represents the sum of reaction rates of the reaction candidates.
$\Delta c_i$ is the concentration change during a single reaction step (the $i$-th step in this case) and is written as:
\begin{equation}
    \Delta c_i = C_i |\Delta M|,
\label{eq:Dc_i}   
\end{equation}
where $C_i$ is the concentration corresponding to a single molecule in the molecular set in our simulation, and $\Delta M$ is the change in the number of molecules in the molecular set during a reaction step.
Using the concentration of the system being simulated (in this study, the impact vapor plume) $P_i/RT_i$ and the total number of molecules in the molecular set $N_{\rm{molecules},i}$, $C_i$ is calculated as:
\begin{equation}
    C_i = \frac{P_i/RT_i}{N_{\rm{molecules},i}}.
\label{eq:C_i}
\end{equation}
In this model, $\Delta M$ is restricted to $\pm 1$ or $\pm 2$ as illustrated in \Figref{fig:ex_reaction}.
For simplicity, when considering the case where $\Delta M\pm 1$, $\Delta c_i$ is directly derived as $C_i$ from Eq.~\eqref{eq:Dc_i}.
From Eqs.~(\ref{eq:kinetic_dJ}) and ~(\ref{eq:Monte_dJ}), $\Delta t_i$, the time step of the $i$-th step, is obtained as follows:
\begin{equation}
    \Delta t_i =\frac{\Delta c_i}{\sum r_i} 
    =\frac{\Delta c_i}{\sum \left(k_{\rm{first},i}C_i+ k_{\rm{second},i}C_i^2\right)} 
    =\frac{1}{\sum k_{\rm{first},i}+ C_i\sum k_{\rm{second},i}} \ \rm s.
\end{equation}
where $k_{\rm{first},i}$ and $k_{\rm{second},i}$ are the rate constants of first-order reactions and second-order reactions, respectively. 
Thus, $\Delta t_i$ is regulated by the reaction rates of the reaction candidates at that step and is dominated by the fastest reaction among the candidates.
%It is also important to mention that $\Delta t_i$ also depends on the temperature and pressure as indicated by Eq.~(\ref{eq:C_i}).

\end{document}